\documentclass[aps,reprint,superscriptaddress,a4paper,prb,showkeys,floatfix]{revtex4-1}

\usepackage{graphicx}
\usepackage{subfigure}  
\usepackage{booktabs}
\usepackage{amssymb}
\usepackage{amsmath}
\usepackage{multirow}
\usepackage{epstopdf}
\usepackage{lineno,hyperref}
\usepackage{color,soul}
\usepackage{tabularx}
\modulolinenumbers[5] 

\begin{document}

\title{Anisotropic three-dimensional weak localization in ultrananocrystalline \\diamond films with nitrogen inclusions}

\author{L.H.~{Willems~van~Beveren}}
\email[Corresponding author: ]{laurensw@unimelb.edu.au}
\author{D.L.~Creedon}
\author{N.~Eikenberg}
\author{K.~Ganesan}
\author{B.C.~Johnson}
\affiliation{School of Physics, University of Melbourne, Parkville, VIC 3010, Australia}
\author{G.~Chimowa}
\altaffiliation{Botswana International University of Science and Technology, \\P. Bag 16, Palapye, Botswana}
\author{D.~Churochkin}
\altaffiliation{Saratov State University, Saratov, 410012 Russia}
\author{S.~Bhattacharyya} 
\altaffiliation{National University of Science and Technology MISIS, \\Moscow, 119049, Leninsky pr. 4, Russia}
\affiliation{Nano-Scale Transport Physics Laboratory, School of Physics, University of the Witwatersrand, Johannesburg 2050, South Africa}
\author{S.~Prawer}
\affiliation{School of Physics, University of Melbourne, Parkville, VIC 3010, Australia}
\date{\today}

\hyphenation{pro-per-ties u-sing du-ring}

\begin{abstract}
We present a study of the structural and electronic properties of ultra-nanocrystalline diamond films that were modified by adding nitrogen to the gas mixture during chemical vapour deposition growth. Hall bar devices were fabricated from the resulting films to investigate their electrical conduction as a function of both temperature and magnetic field. Through low-temperature magnetoresistance measurements, we present strong evidence that the dominant conduction mechanism in these films can be explained by a combination of 3D weak localization (3DWL) and thermally activated hopping at higher temperatures. An anisotropic 3DWL model is then applied to extract the phase-coherence time as function of temperature, which shows evidence of a power law dependence in good agreement with theory.
\end{abstract}


\maketitle

\section{Introduction}
The excellent mechanical, thermal and chemical properties of diamond make it a promising semiconductor for device applications when suitably doped. Ultra-nanocrystalline diamond (UNCD) can be formed on various substrates and the conductivity level and type can be modified by adding a dopant gas during plasma-enhanced chemical vapour deposition growth of the films. For instance, incorporation of boron atoms into the chamber during CVD growth results in films that exhibit $p$-type conduction, and even superconductivity when doped at sufficiently high concentration~\cite{Nesladek:2006}. Similarly, the introduction of sulphur~\cite{PhysRevB.60.R2139} or phosphorus~\cite{KOIZUMI1998540} during the CVD process has been shown to result in $n$-type conduction. Ultra-nanocrystalline diamond has been doped previously by the addition of nitrogen gas during the growth~\cite{Lin:2011}, resulting in a material with conductivity orders of magnitude higher than its intrinsic counterpart. In such nitrogen incorporated UNCD (N-UNCD), $n$-type conduction is observed~\cite{Bhattacharyya:2001}, with a conductivity of as high as $\sigma=143$~S/cm at room temperature. The N-UNCD films consist of grains in both the pure diamond phase (sp$^{3}$-bonded carbon), as well as disordered carbon at the grain boundaries having mixed  sp$^{2}$ and sp$^{3}$ bonding~\cite{Birell:2002}. The exact origin of the $n$-type conductivity in N-UNCD films is not yet fully understood, a knowledge gap that still exists today. It is however believed that the high electrical conductivity is due to a disorder induced change of the carbon bonding configuration~\cite{Zapol:2001}.

Several studies on transport mechanisms in UNCD with $n$-type conduction have been reported. Above \hbox{$>4.2$ K~\cite{Bhattacharyya:2001}}, or \hbox{$>10$ K~\cite{Beloborodov:2006}} hopping or other thermally activated conduction mechanisms dominate. Two-dimensional localization has also been identified to contribute to carrier transport~\cite{Bhattacharyya:2008} in this temperature range. However, little data exists below 4~K and the conduction mechanism in this regime remains an open question. Some previous investigations near 1~K have indicated a dependence of conductivity on sample morphology, with anisotropic weakly localized transport supported by superlattice-like models of the N-UNCD structure~\cite{Shah:2010,Churouchkin:2012}. We show that the morphology has a significant effect on the conductivity, and that studies indicating values of conductivity and an anisotropy parameter (described Section~\ref{TempDepMR})w contradict our present work by a factor of up to two orders of magnitude.

In this work we are able to show that a 3DWL mechanism dominates conduction in the relatively unexplored sub-Kelvin regime. We study the electrical conductivity of Hall bar devices fabricated from N-UNCD films with differing nitrogen content incorporated during the CVD growth, and investigate the temperature dependence of the conductivity and magnetoresistance as low as 36~mK. This represents a significant improvement on previous results. In addition to determining the conduction dependence on the N gas content during growth, two conduction regimes are identified. We confirm that hopping-type conduction dominates at high temperatures, and 3D weak localization dominates in the millikelvin regime. These detailed electrical measurements allow the phase coherence length to be calculated, whose temperature dependence further elucidates the fundamental electrical transport mechanism in these films. The ultra-low temperatures attained in the measurements presented in this work, combined with a detailed theoretical analysis, allow us to present strong evidence for an anisotropic 3DWL model.

\section{Experimental methods}
Electrical measurements were performed on N-UNCD films grown by chemical vapour deposition. These films were deposited on commercially sourced UNCD plates\footnote{Advanced Diamond Technologies Inc., Romeoville, IL, USA. (www.thindiamond.com)}. The electrically insulating UNCD substrates have a typical diamond grain size of 3-5~nm and were grown on conventional oxidized silicon wafers. Through cross-sectional SEM analysis, we measured the oxide (SiO$_{2}$) layer to be $\sim$700~nm thick, with a $\sim$780~nm thick UNCD layer at the surface. An iPlas Cyrannus microwave PECVD reactor (Innovative Plasma Systems GmbH) was used to grow the N-UNCD films using a gas mixture of 1\% CH$_{4}$, $x$\% N$_{2}$, and $(99-x)$\% Ar, where $x$ represents the nitrogen level in the gas feed stock. Samples were fabricated using nitrogen levels of 5\%, 10\%, and 20\%. During growth, the chamber pressure was controlled at approximately 80~Torr, the microwave power at 1250~W, and the substrate temperature at 800$^{\circ}$C. The thickness of the resultant films was measured by SEM to be approximately 940~nm.

For accurate electrical characterization, the as-grown films were then processed into Hall bar shaped device architectures with 400~$\mu$m long channels. The results presented herein were measured on devices having a 10:1 length-to-width ratio in the case of the 10\% and 20\% samples, and a 1:1 length to-width ratio for the 5\% sample. The Hall bars were fabricated using a combination of photolithography and dry etching. An evaporated chromium layer was used as a hard mask to protect the as-grown N-UNCD films during deep reactive ion etching (DRIE). A diamond-compatible DRIE process based on a gas mixture of SF$_{6}$ and O$_{2}$ was used followed by the removal of the mask by wet chemical etching in a mixture of perchloric acid and ceric ammonium nitrate. A second photolithographic step was then used to create a mask for electrical contact evaporation onto the N-UNCD Hall bar. Titanium was evaporated through the shadow mask to a thickness of 20~ nm, followed by 100~nm gold in order to create ohmic contacts. Samples were mounted in a chip carrier and electrical contact was made using aluminium wedge wire bonding.

The electrical characterisation was performed in a cryogen-free dilution refrigerator (CF-450 from Leiden Cryogenics) with a base temperature of 20~mK and equipped with a superconducting vector magnet to provide a magnetic field at an arbitrary angle relative to the film surface. A Keithley 238 source measurement unit (SMU) was used for all electrical measurements together with a low-noise Stanford Research Systems (SRS) 560 voltage pre-amplifier. For transport measurements, a source-drain current $I_{SD}$ is generated by the SMU and passed through the Hall bar channel. The resultant source-drain voltage ($V_{SD}$) is recorded, allowing the calculation of the 2-terminal device resistance $R_{2T}$. To determine the resistivity of the Hall bar channel (avoiding contributions from contacts and other sources of series resistance), the longitudinal voltage drop $V_{xx}$ along the channel is measured using a low-noise voltage pre-amplifier (LNA) in differential input mode, and an Agilent 34410A digital multimeter. The transverse voltage drop across the channel, $V_{xy}$, is measured simultaneously in the same way in order to detect the Hall voltage generated when a magnetic field is applied perpendicular to the substrate. Magnetoresistance measurements were performed with a DC magnetic field applied perpendicular to the Hall bar channel, swept from -6 to +6~T at sufficiently low ramp rates to avoid eddy current induced heating. The magnetic field sweep was also performed in the reverse direction to ensure the data is not subject to magnetic hysteresis or thermal effects. Measurements were performed using DC current-voltage analysis because, unlike AC detection methods, it can reveal non-linear features in the $I$-$V$ curves at low temperatures. These features may be attributed to the formation of Schottky barriers between metallic and semiconducting regions of the device.\\

\subsection{Results}
\subsubsection{Structural measurements}

\begin{figure}[bh]
\includegraphics[width=\linewidth]{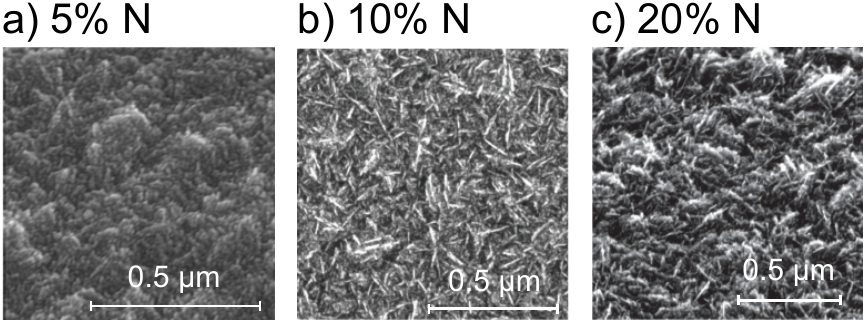}
\caption{\label{fig1} SEM images of the CVD-grown N-UNCD films with different nitrogen content. Subfigures (a) and (c) are taken with a 45 degree tilted substrate.}
\end{figure}

Before turning to the electrical measurements we briefly consider the morphology of the N-UNCD films described above using scanning electron microscopy (SEM) and transmission electron microscopy (TEM), respectively. Typical SEM images of the surface of the 5\%, 10\% and 20\% N-UNCD films are presented in Fig.~\ref{fig1} and show a dependence on the N content used during growth. Compared to the 5\% N-UNCD film which exhibits rounder features, the 10\% and 20\% N-UNCD films appear to consist of sharper or more needle-like features in agreement with previous observations~\cite{Arenal:2007,Vlasov:2007,Tzeng:2014}. In the cross-sectional SEM image, c.f. Fig.~\ref{fig2}(b) and ~\ref{fig2}(c), this coarse morphology is contrasted with the underlying substrate of intrinsic UNCD, which appears smooth due to the resolution limit of the SEM preventing individual nano-scale diamond grains from being distinguished. While the large scale morphology of the 10\% and 20\% samples appears relatively similar, the increase of nitrogen content in the CVD gas mixture is well correlated with an increased fractional volume of the grain boundaries, and an increased fraction of sp$^{2}$ bonded carbon in the grain boundaries~\cite{BirrellBondingStructure}. It is this increase in nano-scale grain boundary thickness and chemical bonding which leads a change in electrical conductivity in UNCD as a function of N content.
Interestingly, in Ref.~\cite{Arenal:2007}, evidence was found for the formation of nanowires with superlattice sub-structure in the plane of the diamond crystal when the concentration of nitrogen was increased in the CVD feed gas. We observe similar structure in our samples, which can be clearly seen in the high-resolution TEM image Fig.~\ref{fig2}(d). The dendritic growth originates from the seed nanodiamonds at the surface of the SiO$_{2}$ layer~\cite{ZOLOTUKHIN201364}, forming vertically oriented needle-like structures that are continuous through the UNCD and N-UNCD layers. From the TEM image Fig.~\ref{fig2}(e) an estimate of the N-UNCD grain size can be made by examining the clear fringes corresponding to grains whose $\{111\}$ lattice planes are parallel to the incident imaging electron beam~\cite{annurev.matsci.29.1.211}. Distinguishable grains are visible against the low contrast background having dimensions of approximately 5~nm on average, a value which is entirely expected based on the size of the seed nanodiamonds used to grow the UNCD film on the SiO$_{2}$ layer.


Despite the presence of N during CVD growth, corresponding changes in the concentration of nitrogen in the diamond bulk were not detected despite extensive further characterisation with Raman, Rutherford backscattering (RBS)~\cite{Ninathesis}, X-ray photoelectron spectroscopy (XPS) and energy dispersive X-ray spectroscopy (EDX) measurements (not shown). Although it has been shown previously that N is incorporated to a small extent in the UNCD film~\cite{Lin:2011}, our data agrees well with secondary ion mass spectrometry (SIMS) and resonant nuclear reaction analysis (RNRA) investigations of N-incorporation in N-UNCD films~\cite{Garratt:2013}. Thus, we conclude from our data that the dominant effect of N incorporation in the gas feed during CVD growth is to change the nanoscale morphology of the UNCD film, rather than the atomic composition of the diamond nanoparticles as such.

Thus, for the following electrical measurements, we assume a 3D conducting material behaving in a bulk-like fashion, where conduction may be a product of the grain boundary character.

\subsubsection{Electronic transport\label{sec:electronictransport}}

\begin{figure}[ht]
\includegraphics[width=\linewidth]{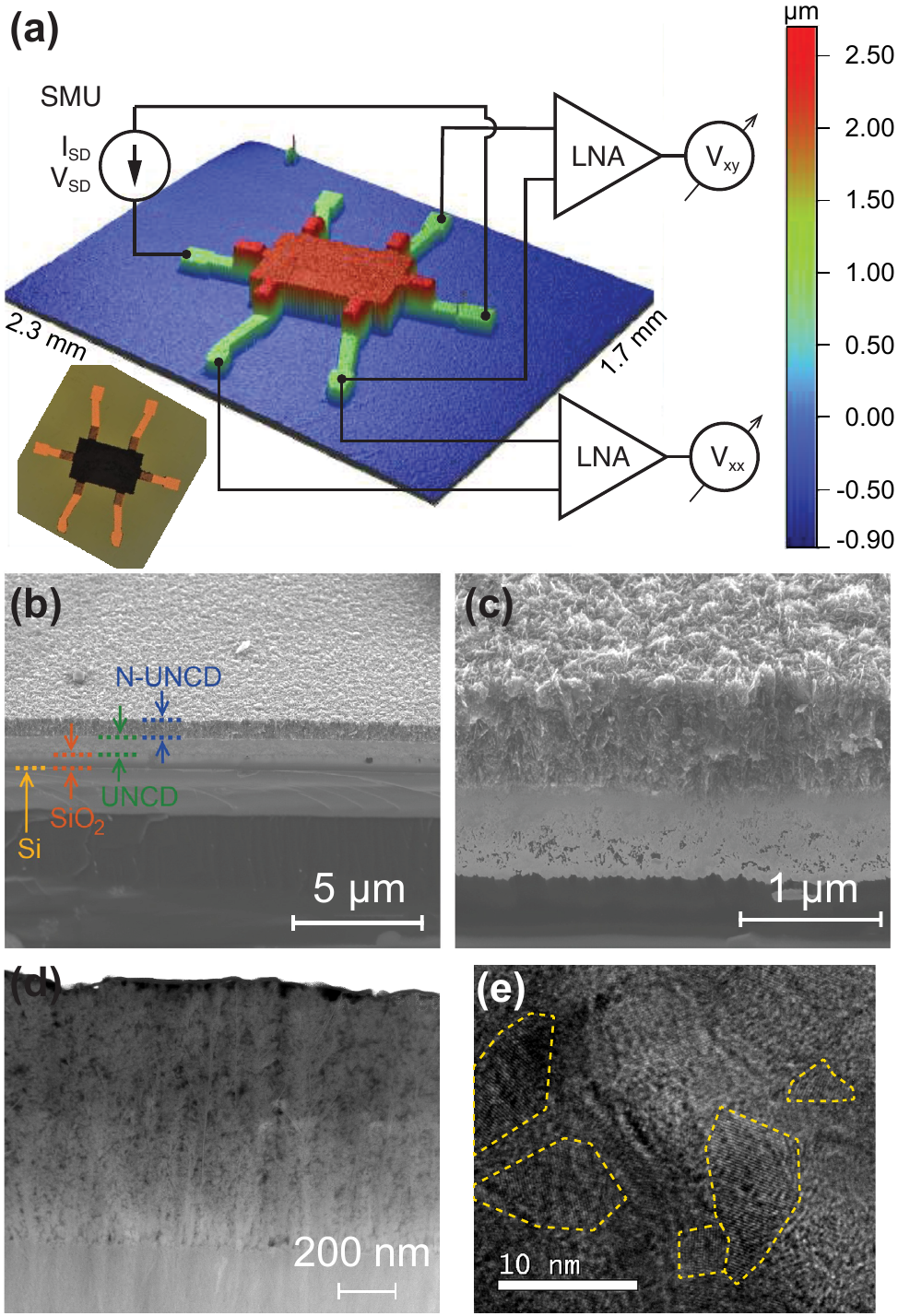}
\caption{\label{fig2} (a) Electrical measurement schematic overlaid on a 3D optical profilometer image of an N-UNCD sample with a channel width of 400~$\mu$m. A constant current is applied through the Hall bar while recording $V_{xx}$ and $V_{xy}$ with voltage pre-amplifiers as function of either temperature or magnetic field (perpendicular to the sample surface). The false-colour scale corresponds to height above the substrate. The colours do not correspond to the layers or regions of the sample. An inset optical micrograph shows the Ti/Au contact regions which overlap the arms of the N-UNCD Hall bar (b) SEM image of a 20\% N-UNCD device in cross-section showing the N-UNCD ($\sim$940 nm), UNCD ($\sim$780 nm), and SiO$_2$ ($\sim$700 nm) layers above the silicon substrate. (c) zoomed cross-sectional image showing the clear difference in morphology between the N-UNCD and UNCD layers. (d) High-angle annular dark-field STEM image of an ion-beam milled cross-section extracted from the 20\% N-UNCD sample, showing a dendritic needle-like growth morphology. (e) TEM image of the 20\% N-UNCD sample with clear individual grains outlined.}
\end{figure}

Figure~\ref{fig2} shows a schematic of the electrical setup used to investigate conduction in these films, along with an 3D optical profilometer image of the 10\% N-UNCD Hall bar. The room temperature I$_{SD}$-V$_{xx}$ traces (taking into account material thickness and Hall bar geometry) were found to be linear (non-rectifying) and for the 5, 10, and 20\% N samples gave a 4-terminal conductivity of 0.01, 1.26 and 11.44~S/cm, respectively. Similar trends for N-UNCD have been reported previously~\cite{Williams:2004,Mares:2006}.

\begin{figure}[!ht]
\includegraphics[width=\linewidth]{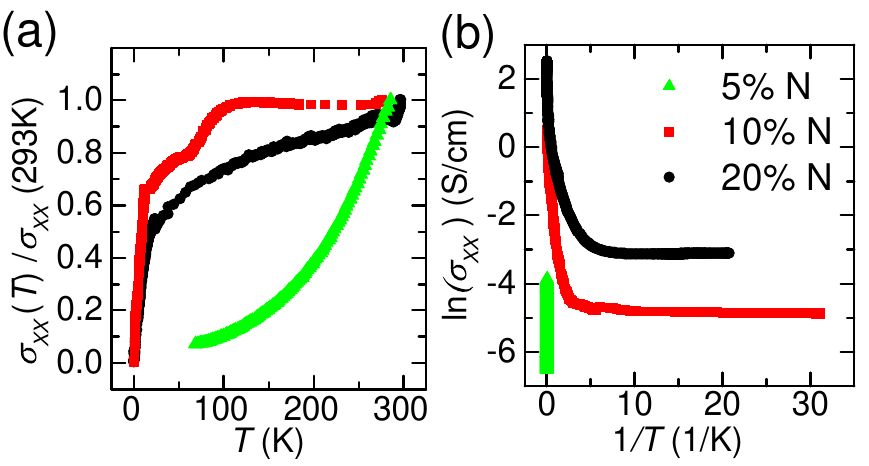}
\caption{\label{fig3} Temperature-dependent conductivity of the N-UNCD films. (a) Conductivity is normalised to the room temperature (293~K) value and plotted as a function of temperature (b) The natural logarithm of the raw conductivity (not normalised) is plotted as a function of reciprocal temperature. A source-drain current of 10~nA was used for the measurement.}
\end{figure}

The temperature dependence of the conductivity for the three samples is shown in Fig.~\ref{fig3}. For all films a drop in the conductivity with decreasing temperature is observed especially below 50~K. The same data set is plot as reciprocal temperature vs. the natural logarithm of conductivity in Fig.~\ref{fig3}(b) to highlight the similarity in the trends for the 10 and 20\% samples. The 5\%~N-UNCD film displays a different dependence on temperature and reduces its conductivity much faster than the other films. Furthermore, the 5\%~N-UNCD film undergoes a carrier freeze out at around 40~K, exhibiting a resistance \hbox{$R_{4T} >  100~$G$\Omega$}, exceeding the limits of our DC measurement capability. In contrast, the 10\%~and 20\% N-UNCD films stay conductive down to the milli-Kelvin regime, which suggests their carrier concentration is above the metal-insulator-transition.

\begin{figure}[tbhp]
\centering
\includegraphics[width=0.8\linewidth]{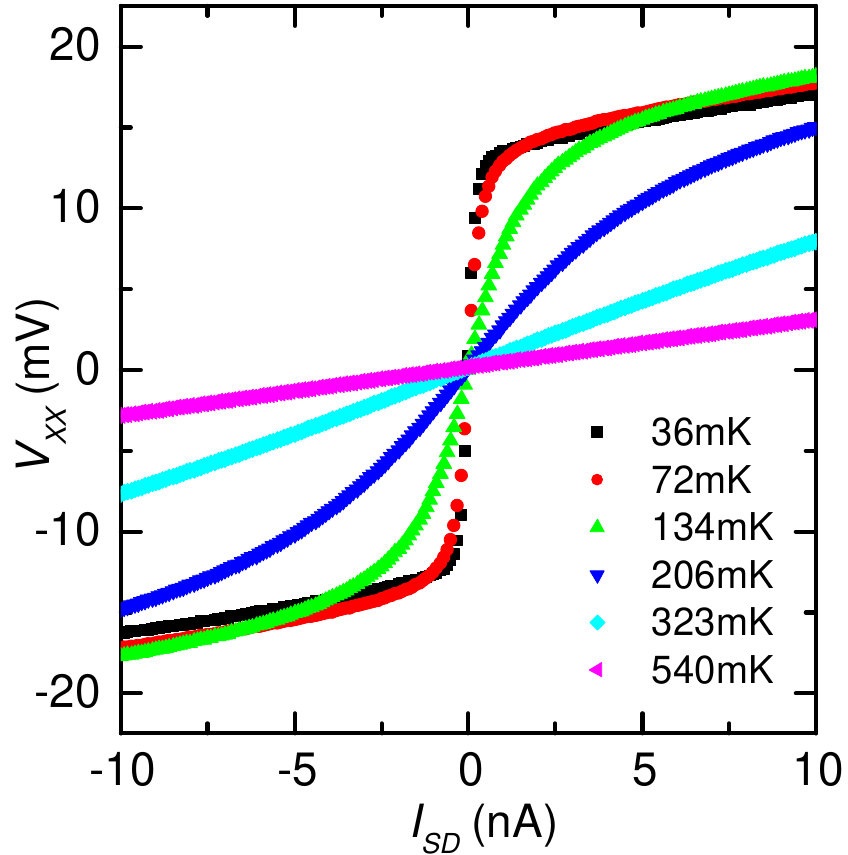}
\caption{\label{fig4} Temperature dependence of the $I$-$V$ curve for the 20\% N-UNCD film. Similar non-linear behaviour was also observed in the 10\% film.}
\end{figure}

For temperatures below approximately 1~K, the otherwise linear $I$-$V$ curves of the 10\%~and 20\% N-UNCD films begin to exhibit unexpected non-linear behaviour in both a 2-terminal and 4-terminal configuration. An example series of $I$-$V$ curves as a function of temperature for the 20\% sample is shown in Fig.~\ref{fig4}. Although the contact resistance determined through current-voltage measurements increases as the sample is cooled, the $I$-$V$ curves remain highly linear and symmetric down to low temperatures. We do not expect the non-linearity that arises is due to a Schottky-like barrier as it is symmetric with respect to bias voltage. The power dissipated in the sample by Joule heating is also negligibly small compared to the cooling power of the cryostat, and thus the behavior cannot be attributed to localised heating effects.

Similar non-linear $I$-$V$ behavior has been seen previously in other disordered systems such as graphene nanoribbons~\cite{Han:2010aa}, carbonized polymer fibers~\cite{Kim:2016aa}, and semiconductor nanocrystal solids~\cite{Yu:2004aa}. The voltage threshold, evident in the $I$-$V$ curve at low temperature, is attributed to a transport gap due to a Coulomb blockade (CB) effect~\cite{GrabertDevoret:1992}, where electron charge localises in a 3D network of insulating islands surrounded by a conducting grain boundary. For this effect to be observed, the temperature must be lower than the charging energy $e^2/2(C_0+9C_i)$, related directly to the self-capacitance $C_0 = 4\pi\epsilon\epsilon_0 r$ of the nanocrystalline islands of radius $r$, and mutual capacitance $C_i = 2\pi\epsilon\epsilon_0 r \ln\left[(r+d)/d\right]$ between islands separated by a distance $2d$~\cite{Joung:2013aa}. The factor of 9 in the denominator arises from the assumption of an average of 9 nearest neighbours per grain~\cite{PhysRevLett.95.156801,Black1131}, and $\epsilon_r$ is the permittivity of the grain boundary or tunnelling barrier.

Applying this Coulomb blockade model to our system, assuming the dielectric constant of diamond $\epsilon_r=5.5$ is valid, and using the average grain size $2r = 5$~nm, and grain separation $2d = 2$~nm determined by TEM (Fig.~\ref{fig2}(e)) and consistent with the literature~\cite{AUCIELLO2010699}, we compute the temperature corresponding to this charging energy to be approximately 80~K. However, the voltage threshold in the $I$-$V$ curve (or transport gap) is only seen to arise at below $\sim$500~mK in the N-UNCD samples.\\

Given the high confidence in the grain and grain boundary dimensions, it is apparent that the assumed value of $\epsilon_r$ is not valid. We provide further evidence and discuss in detail the non-validity of $\epsilon_r$ in Section~\ref{fitting_analysis}, but note here that an effective value of $\epsilon_r = 878$ leads to the experimentally observed Coulomb blockade threshold temperature of $\sim$500~mK. Given the clear energy barrier at low bias, which leads to complicated non-linear $I$-$V$ behavior, we herein only describe results with the Hall bars biased in the linear regime, i.e. $I_{SD}>10$~nA. The current was selected to be as low as possible in the linear regime to prevent resistive (self) heating.

\subsubsection{Low-temperature magnetoresistance}

\begin{figure}[!thbp]
\centering
\includegraphics[width=0.9\linewidth]{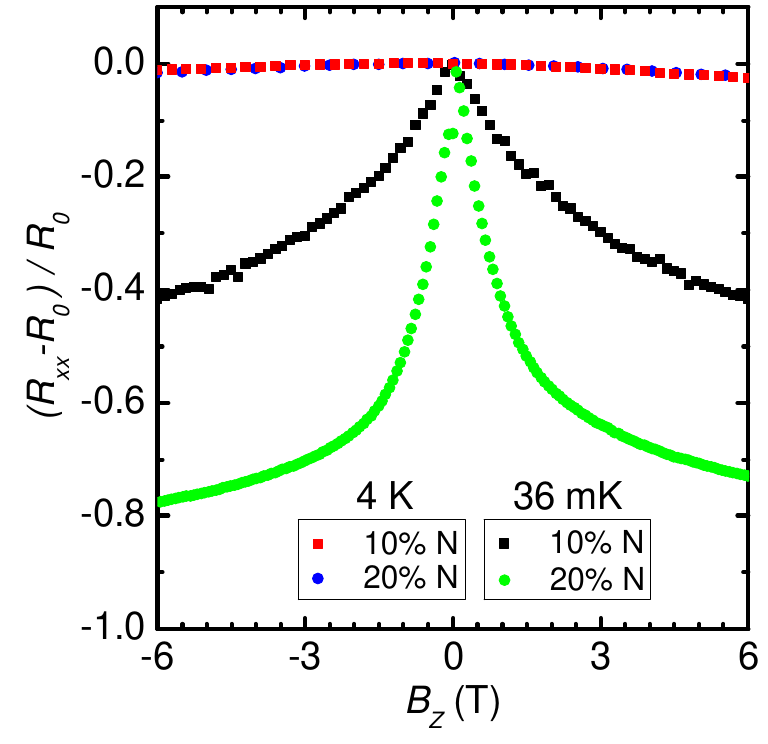}
\caption{\label{fig5} Normalized resistance change as a function of magnetic field for the 10\% (squares) and 20\% (circles) N-UNCD films at $\sim4$~K (blue/red) and 36~mK (black/green). The response of the 10\% sample at 4~K was similar to that of the 20\% N sample.}
\end{figure}

The effect of a perpendicular magnetic field $B_{Z}$ on the 4-terminal resistance $R_{4T}$, i.e. the magnetoresistance, was investigated for the N-UNCD films. The result of these measurements is shown in Fig.~\ref{fig5} comparing the 10\% and 20\% N-UNCD films at two different temperatures.  The results presented used bias currents of 50~nA and 20~nA, respectively, but we did not find any bias dependence of the results when operating above approximately 10~nA, i.e. above the nonlinear region of the IV curve. The magnetoresistance is defined here as \hbox{$(R_{xx}-R_{0})/R_{0}$}, with $R_{xx}=R_{4T}$ the field-dependent resistance, and $R_{0}$ the film resistance with zero applied magnetic field at the temperature specified.

The data show a weak localization (WL) effect in both the 10\% and 20\% N-UNCD films that becomes much stronger with decreasing temperature, as expected by virtue of its quantum mechanical origin. The WL effect is characterized by a negative magnetoresistance (NMR)~\cite{Nesladek:2006b}, resulting in a peak of the (normalized) resistance at zero field (as opposed to a positive magnetoresistance observed in $p$-type diamond~\cite{Willemsvanbeveren:2016} layers). Interestingly, the WL peak is much stronger for films grown with a higher nitrogen concentration. Unlike previous publications, which have only measured MR above 4~K, here we were able to extract the magnetic field dependence at temperatures as low as 36~mK where the MR effects are much more pronounced. This facilitated a detailed analysis of the fundamental transport properties of 10\% and 20\% N-UNCD films. It is worth noting that for such large MR values the magnetoconductance fails to be equal (by approximation) to the inverse of the magnetoresistance. 

As we will show in Section~\ref{section:discussion_analysis}, by fitting the magnetoresistance data with a 3D weak localization model, it is possible to extract the phase coherence length $L_{\phi}$, a parameter characteristic of the strength of this quantum mechanical effect. Furthermore, by repeating this procedure at a range of temperatures we extract the temperature dependence of the phase coherence length, which is predicted to scale as a power law in the case of 3DWL. We then compare the functional dependence of $L_{\phi}$ with the temperature dependence of the conductivity, which is also predicted to follow a power law, to examine how consistently these theoretical models agree.

\section{Discussion and analysis\label{section:discussion_analysis}}
\subsection{Temperature dependent conductivity\label{fitting_analysis}}

\begin{figure}[!bth]
\centering
\includegraphics[width=\linewidth]{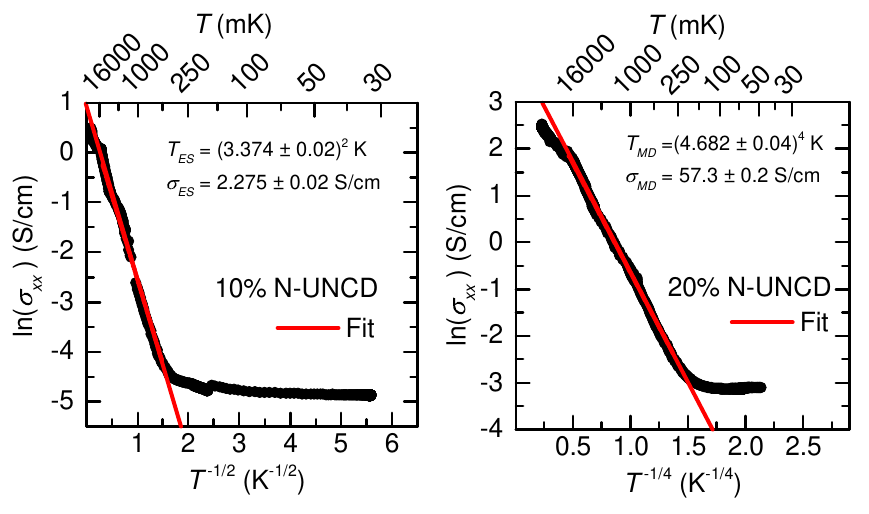}
\caption{\label{fig6} Temperature dependence of the natural logarithm of the conductivity for the 10\ and 20\% N-UNCD samples with fitting functions corresponding to variable range hopping theories. A linear temperature scale is shown on the top x-axis for convenience.}
\end{figure}

The functional dependence of the conductivity on temperature for the 10\% and 20\% N-UNCD samples, is shown in Fig.~\ref{fig6}. In the high-temperature limit, the 10\% N-UNCD sample shows a linear relationship between the natural logarithm of the conductivity and the inverse square root of the temperature. This $T^{-1/2}$ power law corresponds to a dielectric response often seen in granular materials~\cite{Abeles:1975}. Electrical conduction in the dielectric regime of granular metals results from transport of electrons by thermally activated tunnelling (hopping) from one isolated grain to the next. At sub-Kelvin temperatures this type of hopping mechanism is suppressed and a different transport mechanism dominates. Note that this functional temperature dependence is also characteristic of the Efros-Shklovskii (ES) variable range hopping model, which accounts for a Coulomb gap, a small jump in the density of states near the Fermi level due to interactions between localized electrons.

In contrast, the 20\% N-UNCD sample shows a linear relationship between $\ln(\sigma_{xx})$ and $T^{-1/4}$, again indicating variable range hopping (VRH), but this time of the Mott-Davis (MD) kind. However, at sub-Kelvin temperatures this type of hopping is also suppressed and a different transport mechanism becomes apparent.\\


The fitting functions for the 10\% and 20\% N-UNCD samples c.f. Fig.~\ref{fig6} correspond to Eqn.~\ref{eqn:1} and Eqn.~\ref{eqn:2}, respectively:

\begin{align}
	\sigma_{xx}^{(10\%)}&=\sigma_{ES}e^{-(T_{ES}/T)^{1/2}}\label{eqn:1}\\
    \sigma_{xx}^{(20\%)}&=\sigma_{MD}e^{-(T_{MD}/T)^{1/4}}\label{eqn:2}
\end{align}


where $\sigma_{ES}$, $T_{ES}$ and $\sigma_{MD}$, $T_{MD}$ are fitting parameters whose values can be found in Table~\ref{table:1}. The pre-exponential factors $\sigma_{ES}$ and $\sigma_{MD}$ are assumed to be independent of temperature.


Note the subtle difference in the exponent between the two equations. In summary, the 10\% N-UNCD sample exhibits an Efros-Shklovskii~\cite{ShklovskiiEfros:1984} type hopping in the presence of Coulomb interactions~\cite{Beloborodov:2006,Choy:2008}, with \hbox{ln($\sigma_{xx})\propto(T_{ES}/T)^{1/2}$}. In contrast, the hopping mechanism corresponding to the 20\% N-UNCD sample data resembles Mott-Davis type variable range hopping (VRH) in the absence of Coulomb interactions, with \hbox{ln($\sigma_{xx})\propto(T_{MD}/T)^{1/4}$}.\\

From the fits to the experimental data, we extract a characteristic temperature \hbox{$T_{ES}=(3.374\pm0.02)^{2}=11.383$~K} for the 10\% N-UNCD sample, and a conductivity pre-factor of $\sigma_{ES}=2.275\pm0.02$~S/cm. The value of the characteristic temperature $T_{ES}$ along with the dielectric constant of the material $\epsilon_{r}$ can be used to derive an estimate for the localization length $\xi_{ES}$ using Eqn.~\ref{eqn:3}.

\begin{equation}\label{eqn:3}
    \xi_{ES}=\frac{2.8e^2}{4\pi\epsilon_0 \epsilon_r k_B T_{ES}}
\end{equation}


With $T_{ES}=11.383$~K and using the value of relative permittivity for diamond ($\epsilon_{r} = 5.5$), we derive a localization length of $\xi_{ES}=745$~nm. This value is questionably large when compared to the film thickness ($\sim$1~$\mu$m), which may be explained by the choice of dielectric constant. Published values of the dielectric constant of diamond are typically close to what we use here, however values between $\epsilon_{r} = 3.57$--$w.5$ have been reported~\cite{TF9403500575}. It should be noted however that $\epsilon_{r}$ for nitrogen incorporated UNCD is expected to be different from intrinsic diamond in a bulk sense, perhaps significantly so, as a relatively large volume fraction of the material is made up of the grain boundaries where electric transport occurs. These grain boundaries incorporate a high density of defects, conductive non-diamond carbon phases, and other impurities such as trans-polyacetylene. Given that conduction does not occur in the intrinsic diamond grains, and instead wholly in the grain boundaries, it is clear that an effective dielectric constant for this region must be used. Values for the dielectric constant of N-UNCD in the literature are not commonly found, though values for microcrystalline diamond as high as $\epsilon_{r} = 60.3$, and nanocrystalline diamond as high as $\epsilon_{r} = 3003$ have been reported~\cite{epsilonUNCD}. In addition, it has been shown previously that granular materials exhibiting hopping conduction in the vicinity of a metal-to-insulator transition undergo a divergence of the dielectric constant~\cite{Entin_Wohlman_1983}. By instead fixing the localization length $\xi_{ES}$ to the average N-UNCD grain size determined by TEM analysis to be approximately 5~nm (see Fig.~\ref{fig2}(e)), Eqn.~\ref{eqn:3} gives an effective dielectric constant of $\epsilon_r = 822$, which is within the range seen previously in nanocrystalline diamond, and consistent with the value for $\epsilon_r$ derived from our Coulomb blockade analysis of the $I$-$V$ behavior of the sample. Alternatively, taking the value of $\epsilon_r = 879$ derived from the CB analysis, a localization length of $\xi_{ES}=4.68$~nm is computed, which closely matches the measured average grain size and thus the expected localization length. We therefore consider this compelling evidence that the effective dielectric constant in the N-UNCD grain boundaries diverges considerably from the value of intrinsic diamond at the extreme of low temperature.

The Efros-Shklovskii VRH model also provides an estimate of the (temperature and $\xi_{ES}$ dependent) average hopping distance, and average hopping energy:

\begin{align}
    \overline{R}^{ES}_{hop} &=\frac{1}{4} \xi_{ES} \left(\frac{T_{ES}}{T}\right)^{1/2} \label{eqn:4}\\
	\overline{W}^{ES}_{hop} &=\frac{1}{2} k_B T \left(\frac{T_{ES}}{T}\right)^{1/2} \label{eqn:5}
\end{align}

Using the localization length of $4.68$~nm, these result in an average hopping distance \hbox{$\overline{R}^{ES}_{hop}=3.22$~nm}, and energy \hbox{$\overline{W}^{ES}_{hop}=145~\mu$eV} at a temperature of 1~K. Furthermore, a value for the Coulomb gap width $\Delta_{CG}$ can be obtained using Eqn.~\ref{eqn:6}:

\begin{equation}\label{eqn:6}
 	\Delta_{CG}	=\frac{e^3\sqrt{N(E_{F})}}{\epsilon_{r}^{3/2}}
\end{equation}

Another estimate of the Coulomb gap width can be obtained independent of the density of states and relative dielectric constant with:

\begin{equation}\label{eqn:7}
     \Delta_{CG}\approx k_B\left(\frac{T_{ES}^3}{T_{MD}}\right)^{1/2}
\end{equation}

Using Eqn.~\ref{eqn:7} results in a value for the Coulomb gap of $\Delta_{CG}=151.57~\mu$eV.\\

From the fit to the temperature dependent conductivity data for the 20\% N-UNCD sample, we extract a characteristic temperature $T_{MD}=(4.682\pm0.04)^{4}=478.0$~K and a conductivity pre-factor of $\sigma_{MD}=57.3\pm0.2$~S/cm. The characteristic temperature $T_{MD}$ can be used to derive an estimate for the density of states at the Fermi level $N(E_{F})$, after fixing a value for the localization length $\xi_{MD}$. When we fix the value of the localization length to match the typical N-UNCD grain size of $\xi_{MD}=5$~nm, we find a density of states at the Fermi level of $N(E_{F})=2.18\times10^{46}$~J$^{-1}$m$^{-3}$ or $3.49\times10^{21}$~eV$^{-1}$cm$^{-3}$. From Hall measurements, using a sensitive AC lock-in amplifier technique, we estimate an electron carrier density of $n_{3D}\approx1.1\times10^{19}$~cm$^{-3}$ in this material at 0.2~K together with an associated carrier mobility of $\mu\approx1.8$~cm$^{2}$/V$\cdot$s, resulting in an elastic scattering time $\tau_{0}=\mu m^{*}/e=0.57$~ps, with $m^{*}=(1/18)\cdot m_{e}$.\\

Additionally, an estimate of the average Mott hopping distance and average Mott hopping energy can be made within the Mott-Davis VRH theory (which is again temperature and $\xi_{MD}$ dependent) using Eqn.~\ref{eqn:8} and Eqn.~\ref{eqn:9}:

\begin{align}
    \overline{R}^{MD}_{hop} &=\frac{3}{8} \xi_{MD} \left(\frac{T_{MD}}{T}\right)^{1/4} \label{eqn:8}\\
     \overline{W}^{MD}_{hop} &=\frac{1}{4} k_B T \left(\frac{T_{MD}}{T}\right)^{1/4} \label{eqn:9}
\end{align}

This results in values of \hbox{$\overline{R}^{MD}_{hop}=8.76$~nm} and \hbox{$\overline{W}^{MD}_{hop}=100.87$~$\mu$eV} at 1~K.\\


In the low-temperature regime however, we expect that Anderson localization -- possibly in combination with electron-electron interactions -- may dominate electronic transport, with thermally activated hopping mechanisms being suppressed. Since we are dealing with films much thicker than the elastic mean free path, we consider bulk-like transport in the N-UNCD devices, rather than a (quasi) two-dimensional type of transport similar to that reported previously~\cite{Bhattacharyya:2008}. Therefore, in the low temperature limit we employ a 3D weak localization (3DWL) model to fit our data following Eqn.~\ref{eqn:10}:

\begin{equation}\label{eqn:10}
    \sigma_{xx}=\sigma_{0} + \frac{e^2}{\hbar \pi^3} \frac{1}{a} T^{p/2}
\end{equation}

\begin{figure}[th]
\includegraphics[width=\linewidth]{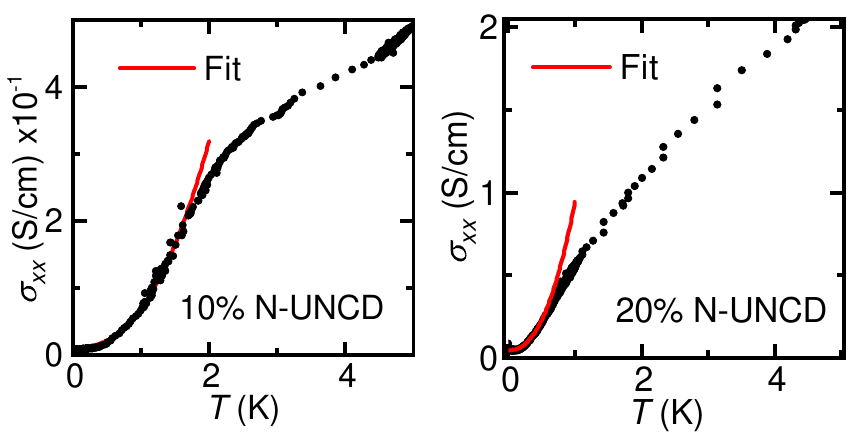}%
\caption{\label{fig7} Temperature dependent conductivity for the 10\% and 20\% N-UNCD films in the Kelvin regime indicating that 3D weak localization theory c.f. Eqn.~\ref{eqn:10} fits the data well. The bias current $I_{SD}$ was 10~nA in both measurements.}
\end{figure}

\begin{table*}[!hbt]
\begin{tabularx}{\textwidth}{lrrrr}
\hline\hline
Sample                  & Fitting method                        & \multicolumn{3}{c}{\hspace{4pc}Fitting parameters} \\ \hline\hline
\multirow{4}{*}{10\% N-UNCD\hspace{4pc}} & \multirow{2}{*}{Efros-Shklovskii VRH} & \hspace{4pc}$T_{ES}$ [K] 	& \hspace{4pc}$\sigma_{ES}$ [S/cm] 	& \hspace{4pc}$\xi_{ES}$ [nm] 						\\
                        &                                       & 11.383 		& 2.28 					& 4.68  								\\ \cline{2-5}
                        & \multirow{2}{*}{3DWL}                 & $p$ 			& $a$ [nm] 				& $\sigma_{0}$ [S/cm] 					\\
                        &                                       & 4.02 			& 1187.00 				& 0.071 								\\ \cline{1-5}
\multirow{4}{*}{20\% N-UNCD\hspace{4pc}} & \multirow{2}{*}{Mott-Davis VRH}       & $T_{MD}$ [K] 	& $\sigma_{MD}$ [S/cm] 	& $N(E_{F}$) [eV$^{-1}$cm$^{-3}$] 	\\
                        &                                       & 478.00 		& 57.30 				& 3.49$\times$10$^{21}$  				\\ \cline{2-5}
                        & \multirow{2}{*}{3DWL}                 & $p$ 			& $a$ [nm] 				& $\sigma_{0}$ [S/cm]         			\\
                        &                                       & 4.75 			& 89.22 				& 0.042\\
\hline\hline
\end{tabularx}
\caption{\label{table:1} Fitting parameters extracted from VRH and 3DWL theories for both 10\% and 20\% N-UNCD films. Note that the value for $\xi_{ES}$ extracted from the Efros-Shklovskii fitting parameters relies strongly on the value chosen for the dielectric constant, discussed in Section~\ref{fitting_analysis}.}
\end{table*}

with $\sigma_{0}$, $a$ and $p$ the fitting parameters. Here, $\sigma_{0}$ is again assumed a temperature independent constant, $a$ represents a localization length and $p$ determines the exponent of the power law and comes from the scaling theory of localization and Coulomb interaction model~\cite{Lee:1985}. It should be noted that this exponent makes a difference in the temperature dependence of the weak localization effect ($T^{p/2}$) for the 3D case. For 2D systems, the temperature dependence ln$(T)$ is always the same, i.e. independent of $p$. The theoretically predicted values for $p$ can be found in Ref.~\cite{Lee:1985} and are given as $p=1.5$, 2 or 3, depending on whether electron-electron scattering is in the dirty vs. clean limit or whether electron-phonon scattering dominates the inelastic scattering rate, respectively. Furthermore, the inelastic scattering time, defined as $\tau_{\phi}\propto T^{-p}$, $p>1$, directly determines the Thouless length~\cite{Thouless:1977} through $L_{Th}=\sqrt{D \tau_{\phi}}$, with $D$ the diffusion constant. The Thouless length represents the mean distance between successive inelastic scattering events suffered by a charged particle in a diffusive conductor. Or stated differently, the Thouless length represents the distance over which a charged particle travels without losing its quantum-mechanical phase information, responsible for the weak localization effect through self interference.

In this work, we refer to this Thouless length scale as $L_{\phi}$ and it is our aim to determine the value of $p$ from electrical (magneto)transport measurements. The extraction of the value of $p$ allows us to identify whether the electrical conduction at low-temperature is indeed 3D and if it is dominated by localization as opposed to electronic Coulomb interactions~\cite{Wang:2006}. Finally, it allows us to compare the electronic transport properties of our N-UNCD films with results from other researchers. In general, for 3DWL and a material that can be treated as a bulk dirty metal (characterized by $k_{B}T<\hbar/\tau_{0}$, where $\tau_{0}$ is the elastic scattering time), the parameter $p>1$ in Eqn.~\ref{eqn:10}.

We fitted our conductivity data of the 10\% and 20\% N-UNCD films in the sub--4 K temperature regime according to the available theories dealing with Anderson localization, electron-electron interactions and electronic screening functions outlined in Refs.~\cite{Du:1998,Altshuler:1985,Kobayashi:1985,Shah:2010} (see Fig.~\ref{fig7}). Since the $e$-$e$ interaction term is independent of magnetic field but the weak localization term is not~\cite{Altshuler:1985}, performing magnetoresistance measurements enables one to directly distinguish $e$-$e$ interaction from weak localization and unravel the dominant transport mechanism.

In Fig.~\ref{fig7} the zero-field conductivity of the N-UNCD films is fitted based on Eqn.~\ref{eqn:10}, which originates in 3DWL theory~\cite{Lee:1985} and assumes \textit{a priori} no explicit value of the exponent in the temperature power law. We find that in the sub-Kelvin temperature range, the conductivity data for both 10\% and 20\% N-UNCD films is well modelled by the 3D weak localization theory. The resultant 3DWL fitting parameters for both films are summarized in Table~\ref{table:1}. Note that the value of $p$ in the exponent of the temperature power law is $p=4.0-4.8$, which is somewhat larger than the previously mentioned estimates for $p$ in the literature based on dephasing mechanism and dimensionality. However an exponent $p=4$ has been seen previously in germanium nanowires~\cite{Sett_2017} and disordered metals~\cite{PhysRevB.60.3940} and is attributed to 3DWL with electron-phonon interaction in the presence of strong impurity scattering in the dirty limit~\cite{Lin_2002}.

Therefore, from the MR traces and conductivity data we come to the conclusion that in the temperature regime below 1~K, the 3DWL mechanism dominates the electronic transport over, but not excluding, $e$-$e$ interactions. In the higher temperature range (see Fig.~\ref{fig7}) the conductivity data deviates from the 3DWL fit as thermally activated VRH conduction starts to overtake. In this VRH regime, it is in principle possible to independently determine a localization length parameter by performing temperature-dependent resistivity measurements at a constant magnetic field~\cite{Ghosh:1998,ShklovskiiEfros:1984,Zabrodskii:1998}. To further confirm the 3DWL nature of the transport mechanism in our N-UNCD films, we performed weak localization measurements with the magnetic field applied perpendicular to the sample and compared it to weak localization measurements in the case where the field was applied parallel to the sample surface using a superconducting 3D vector magnet. As expected for a truly 3D material, the WL data (not shown) of these two distinct field-sample configurations gave identical results.

\subsection{Temperature dependence magnetoresistance}
\label{TempDepMR}
Magnetotransport measurements on N-UNCD and B-NCD films have been reported previously, and it was shown that weak localization effects play an important role~\cite{Williams:2004}. The theory of weak localization for 3D transport is different than for the 2D case~\cite{Hikami:1980,Bergmann:1984,Akhgar:2016}. To reiterate, since the thickness of the conducting films is much larger than both the elastic mean free path and the phase coherence length of electrons (even down to the lowest temperatures) in the diffusive N-UNCD material, we assume the 3D case and therefore 3DWL theory.

The theory of (anisotropic) 3DWL was established by Kawabata~\cite{Kawabata:1980}. This theory was then extended by Bryksin and Kleinert~\cite{BKtheory:1996} (B-K theory) to include the angular dependence of electronic transport, where the magnetic field can have an arbitrary orientation with respect to a set of superlattice (SL) planes~\cite{Gougam:1999}. The B-K theory has already been successfully applied to explain strong anisotropic WL effects in N-UNCD materials~\cite{Bhattacharyya:2014}, based on the idea of a disordered superlattice in the conducting regions of the material.

The motivation for the formation of a SL is based on the work of Ref.~\cite{Vlasov:2012}, where a nitriding-induced rearrangement of the UNCD structure takes place, leading to the formation of diamond nanorods with a graphite shell. The authors claim that in highly nitrided films, it is most probable that either rods gradually form two-dimensional arrays or favorable conditions appear for the formation of nanodisk-shaped carbon structures. Similar structures (nanowires and standing carbon platelets) were also reported in Ref.~\cite{Tzeng:2014}. A recent paper also observed vertically oriented ribbon-like carbon nanowall structures in N-UNCD~\cite{Hejazi:2019aa}. Indeed, the cross-sectional TEM results in the present work (Fig.\ref{fig2}(d)) show clear evidence of vertically oriented dendritic or needle-like structure, leading naturally to the consideration of anisotropic conduction in the material.\\

Here, we present exactly that proposed model for the conduction in our N-UNCD films, in the context of a disordered superlattice-like structure~\cite{Chimowa:2012} in the grain boundary. We further attempt to validate this reported 3DWL anisotropic model, used to explain the conductivity in our N-UNCD films, by investigating the temperature dependence of the phase coherence length~\cite{Shah:2010,Churouchkin:2012}.

At first, the 4~K MR data of the 10\% and 20\% doped N-UNCD films shown (blue/red) in Fig.~\ref{fig5} were fitted applying the 3D anisotropic WL approach based on a propagative Fermi surface (PFS) model. The PFS model was originally developed to explain transport in disordered artificial superlattices. In contrast, an anisotropic 3DWL approach in the diffusive Fermi surface (DFS) limit was recently used to explain the transport in nanocrystalline silicon films~\cite{Pusep:2005} to account for an artificially formed SL structure~\cite{Chiquito:2002,Pusep:2003}. This approach~\cite{Chenai:1995} yields the following basic expression for the conductivity (resistance) correction:

\begin{equation}
\frac{\Delta\sigma(B)}{\sigma_{0}}=-\frac{\Delta R(B)}{R_{0}}= \frac{\alpha}{\sigma_{0}}\frac{e^{2}}{2\pi^2\hbar}\sqrt{\frac{eB}{\hbar}}f_{3}(x)
\label{anisotropic3DWL}
\end{equation}

Here, $x=\frac{\hbar/eB}{4D_{||}\tau_{\phi}}$, $\sigma_{0}$ equals $\sigma(0,T)$, and $R_{0}$ equals $R(0,T)$, i.e. the temperature dependent conductivity (resistance) at zero magnetic field, $\alpha=\sqrt{\frac{D_{||}}{D_{\perp}}}$ is the anisotropy parameter, relating the diffusion constant parallel and perpendicular to the normal of the SL. We also define the Kawabata function $f_{3}(x)$:

\begin{multline}
f_{3}(x)=\sum_{N=0}^\infty \left(2[(N+1+x)^{1/2}-(N+x)^{1/2}]\right.\\
\left.-(N+\frac{1}{2}+x)^{-1/2}\right)
\label{Kawabata}
\end{multline}

where $N$ is the Landau quantum number, i.e. the discrete energy levels occupied by cyclotron orbits of the electrons in a magnetic field. For the fit we chose $N=100$ and $\sigma(0,T)=57.3$~S/m at $T=4$~K for the 10\% doped N-UNCD film.

However, the only way we could match the MR data with the fitting function in Eqn.~\ref{anisotropic3DWL} is by fixing the anisotropy factor to $\alpha=0.01$ (0.048 for the 20\% doped N-UNCD film). The smallness of the parameter $\alpha$ is entirely determined by the unexpected smallness of the parameter $\sigma_{0}=\sigma(0,T)$. We find that $\sigma_{0}=2.8$~S/m at $T$=36~mK, which is much lower than previous measurements on this material~\cite{Shah:2010} at sub-Kelvin temperatures, where $\sigma_{0}$ was measured to be 1500~S/m. This difference in conductivity corresponds to a difference in $\alpha$ by a factor of $\sim500$, giving a value for $\alpha$ of 1--5 for previous results~\cite{Chimowa:2012}. The discrepancy may be explained by the morphology of the assumed layers in the 10\% and 20\% doped N-UNCD films, which suggests that the assumed SL layers were perpendicular to the current direction. In other words, the interlayer (tunneling) current and its corresponding conductivity were measured.

\begin{figure}[t!]
\centering
\includegraphics[width=0.9\linewidth]{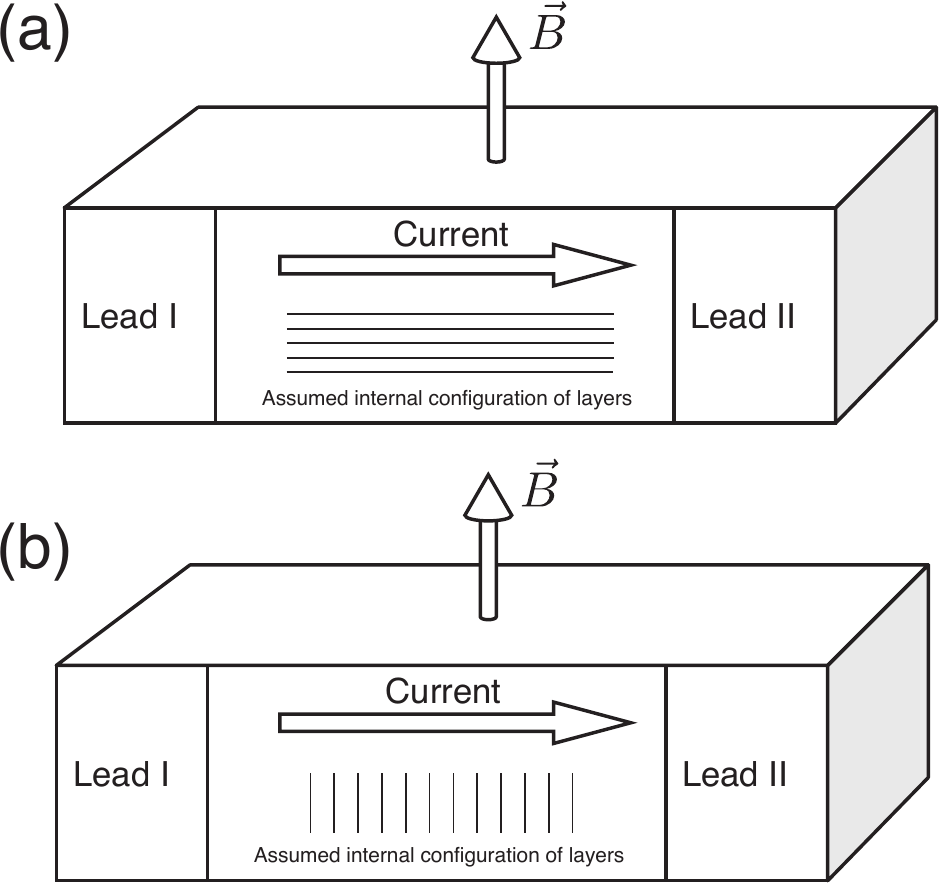}
\caption{\label{fig8} Case I: B-field is aligned parallel to the normal of the superlattice planes. (b) Case II: B-field is aligned perpendicular to the normal of the superlattice planes.}
\end{figure}

Typically for an anisotropic model, the in-layer conductivity is much larger than the interlayer one with $\alpha>1$. But having applied the Kawabata model to our data assuming the `Case I' superlattice configuration shown schematically in Fig.~\ref{fig8}(a), it was revealed that the proper fit can be achieved only if we take the value of the anisotropy parameter around $\alpha=$~0.01, which contradicts the superlattice concept, leading to the conclusion that the in-layer conductivity becomes less than the interlayer one.

\begin{figure}[b!]
\centering
\includegraphics[width=\linewidth]{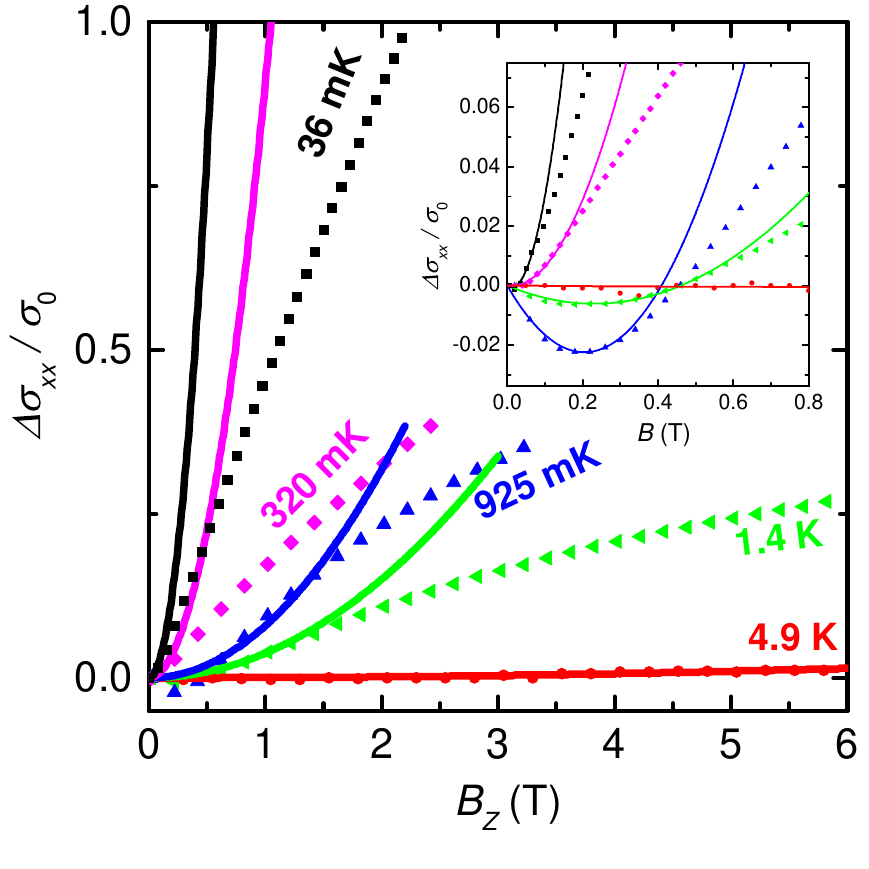}
\caption{\label{fig9} Temperature dependent magnetoresistance data (points) for the 10\% N-UNCD sample, and corresponding \hbox{B-K} fitting functions (solid lines) for $T=$~4.9 (red), 1.4 (green), 0.925 (blue), 0.32 (magenta) and 36~mK (black). Inset: Enlarged portion of the main plot that more clearly demonstrates, by fitting to a second-order polynomial, the deviation from quadratic behaviour past a very low threshold of magnetic field.}
\end{figure}

%
%

Therefore, we focused on the SL configuration of Case II (Fig.~\ref{fig8}(b)), which is nothing but the claim that the measured conductivity was the interlayer one. The Kawabata model should not be applicable here because it was developed solely for the fixed (perpendicular) direction of the magnetic field in respect to the configuration of the layers. From the definition of the conductivity tensor components it follows that $\sigma_{||}=\alpha^{2}$$\sigma_{\perp}$.

\begin{figure}[t!]
\centering
\includegraphics[width=0.9\linewidth]{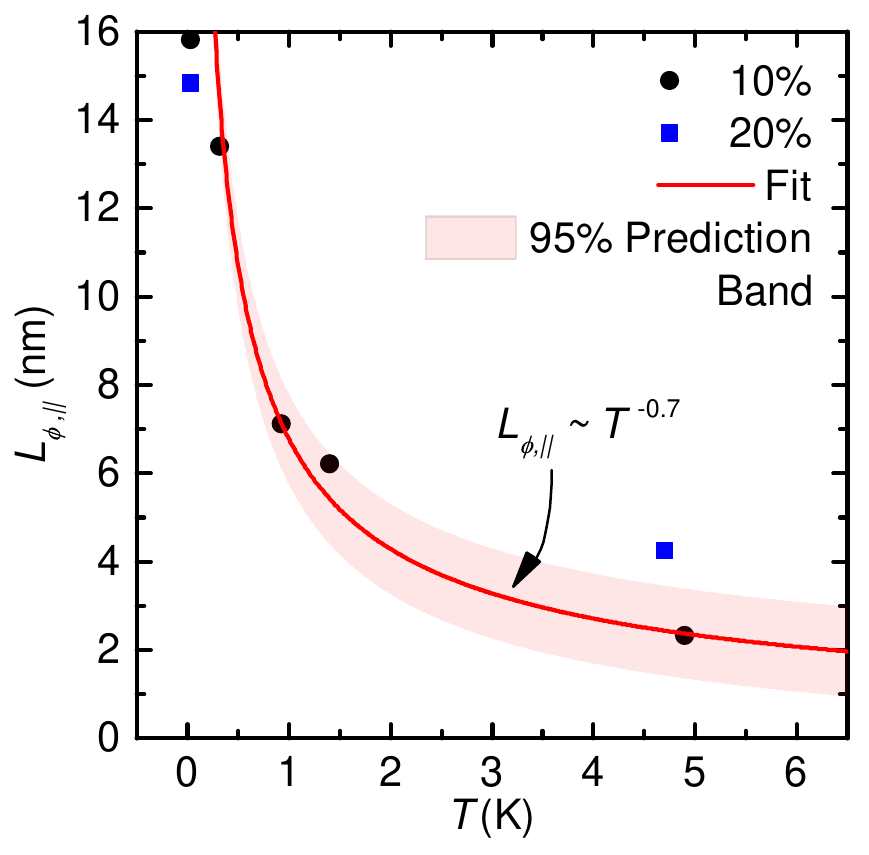}
\caption{\label{fig10} Temperature dependence of the phase coherence length $L_{\phi,||}$ of the 10\% N-UNCD sample. The data is fitted with a power law $T^{p}$ whose exponent is $p\approx-0.7$.}
\end{figure}

Interestingly, for a SL configuration of Case I, both models (either 3D anisotropic WL or B-K theory) can be applied without loss of generality and the expected result should be the same. However, we have strong evidence that the present samples exhibit the SL configuration of Case II, which is consistent with the cross-sectional TEM micrograph (Fig.~\ref{fig2}(d)) showing vertically oriented dendritic nanowire type structure. Therefore we accordingly consider the interlayer component of the conductivity under the condition that the magnetic field is directed along the layers. Having this in mind we utilised the advantages of the B-K approach on the base of which the arbitrary mutual layer-field configurations can be treated. Note that the standard WL approach is a perturbation approach, i.e. the most reliable fit should be expected in the low magnetic field region. As a consequence of that, we made the asymptotic B-K fit for the SL configuration of Case II and show the result of the fitting for five different temperatures in Fig.~\ref{fig9}. Note that the asymptotic B-K fitting formula transforms from Eqn.~20 ($\Delta\sigma_{\perp}(B)$) in the B-K paper~\cite{BKtheory:1996} to the following equation in case of $\Delta\sigma_{||}(B)$:

\begin{equation}
\frac{\Delta\sigma_{||}(B)}{\sigma_{0}}=\frac{e^{2}}{\pi^2\hbar}\frac{L_{\phi,||}^{3}}{\sigma_{0} 24 L(B)^{4}}
\label{DeltaSigmaPar}
\end{equation}
where $L(B)=\sqrt{\hbar/eB}$ is the magnetic length and \hbox{$L_{\phi,||}=\sqrt{D_{||}\tau_{\phi}}$} is the phase coherence length parallel to the normal of the layers, i.e. along the current direction.

The result of the low-magnetic field approximation in the B-K theory is that the normalized conductivity change $\Delta\sigma_{||}(B)/\sigma_{0}$ follows a quadratic behaviour in the low-field limit. This implies that the fitting of the MR data in the high-field regime is by definition poor. Note that this is especially the case for the MR traces at the lowest temperatures, where the field-induced conductivity change is largest. Unlike anisotropic 3DWL (Kawabata) theory, where the high-field limit data can be approximated with a square root behaviour, we are left with extracting the phase coherence length from the fit to the data points in the low-field regime.  Nevertheless, the high-field MR does follow a square root dependence on the magnetic field.  Even considering these several factors, we must stress that no model appears to fit the experimental data with a great deal of certainty, and therefore it is entirely possible that more complex physics is yet to be elucidated in this system. Presently we are not aware of any magnetoresistance theory that can account for the behaviour of these N-UNCD films, other than the anisotropic superlattice model in BK theory. We have therefore applied that model here, with stated assumptions and caveats.




In Fig.~\ref{fig10}, we plot the extracted value of $L_{\phi,||}$ at each temperature. The data show that $L_{\phi,||}$ scales as $T^{-0.7}$ which agrees reasonably well with 3DWL theory and reported values in the literature. In a subsequent publication we will report on recently observed resistivity anisotropy~\cite{Sagar:2015a} of our N-UNCD films at low temperatures when performing angular-dependent magnetic field rotations (at fixed magnitude) using our 3D vector magnet.\\


\section{Conclusion}
We have fabricated N-UNCD films of varying nitrogen content via CVD, with SEM imaging revealing that the morphology of these films changes as a direct result of this variation. We did not detect incorporation of nitrogen into the diamond grains within the limits of further XPS, RBS, and Raman spectroscopy measurements. The electrical conductivity in these films increases with N content, most likely due to an increase in sp$^{2}$ bonds within the grain boundaries. Electrical transport measurements show evidence for a regime of hopping conduction, and an anisotropic 3DWL regime at sub-Kelvin temperatures. Non-linear $I$-$V$ curves below 0.5~K show evidence of a potential energy barrier for the transport in the N-UNCD material, where hopping transport is suppressed. Magnetoresistance measurements as a function of temperature revealed a power law for the phase coherence length, whose exponent agrees with the literature, confirming that the 3DWL model is indeed the suitable model to describe electronic transport in this material at such low temperatures.

It is clear that further work is required to fully and accurately characterise the morphology dependence of the transport mechanism in N-UNCD in the limit of temperature approaching absolute zero. A greater understanding may lead to a host of applications for N-UNCD devices which could exploit the quantum nature of, for instance, the spin associated with weak localization orbits. These films could also find application in low-cost magnetic field sensors based on their anisotropic magnetoresistance properties~\cite{Sagar:2015b} or as electron emitters through field-emission~\cite{Yuan:2016}. Finally, improved understanding of the fundamental electrical properties of N-UNCD films may be also helpful in the development of implantable electronic devices~\cite{Ganesan:2014}.

\section*{Acknowledgments}
This work was performed in part at the Melbourne Centre for Nanofabrication (MCN) in the Victorian Node of the Australian National Fabrication Facility (ANFF). The authors would like to acknowledge A.R. Hamilton for fruitful discussions and S.~Rubanov and K.~Fox for assisting with the material characterization. S.B. also acknowledges financial support from the Ministry of Education and Science of the Russian Federation in the framework of the Increased Competitiveness Program of NUST (MISIS) (grant No.~K3-2018-043).

\bibliography{NUNCDbibliography}

\begin{thebibliography}{67}%
\makeatletter
\providecommand \@ifxundefined [1]{%
 \@ifx{#1\undefined}
}%
\providecommand \@ifnum [1]{%
 \ifnum #1\expandafter \@firstoftwo
 \else \expandafter \@secondoftwo
 \fi
}%
\providecommand \@ifx [1]{%
 \ifx #1\expandafter \@firstoftwo
 \else \expandafter \@secondoftwo
 \fi
}%
\providecommand \natexlab [1]{#1}%
\providecommand \enquote  [1]{``#1''}%
\providecommand \bibnamefont  [1]{#1}%
\providecommand \bibfnamefont [1]{#1}%
\providecommand \citenamefont [1]{#1}%
\providecommand \href@noop [0]{\@secondoftwo}%
\providecommand \href [0]{\begingroup \@sanitize@url \@href}%
\providecommand \@href[1]{\@@startlink{#1}\@@href}%
\providecommand \@@href[1]{\endgroup#1\@@endlink}%
\providecommand \@sanitize@url [0]{\catcode `\\12\catcode `\$12\catcode
  `\&12\catcode `\#12\catcode `\^12\catcode `\_12\catcode `\%12\relax}%
\providecommand \@@startlink[1]{}%
\providecommand \@@endlink[0]{}%
\providecommand \url  [0]{\begingroup\@sanitize@url \@url }%
\providecommand \@url [1]{\endgroup\@href {#1}{\urlprefix }}%
\providecommand \urlprefix  [0]{URL }%
\providecommand \Eprint [0]{\href }%
\providecommand \doibase [0]{https://doi.org/}%
\providecommand \selectlanguage [0]{\@gobble}%
\providecommand \bibinfo  [0]{\@secondoftwo}%
\providecommand \bibfield  [0]{\@secondoftwo}%
\providecommand \translation [1]{[#1]}%
\providecommand \BibitemOpen [0]{}%
\providecommand \bibitemStop [0]{}%
\providecommand \bibitemNoStop [0]{.\EOS\space}%
\providecommand \EOS [0]{\spacefactor3000\relax}%
\providecommand \BibitemShut  [1]{\csname bibitem#1\endcsname}%
\let\auto@bib@innerbib\@empty
\bibitem [{\citenamefont {Nesl\'{a}dek}\ \emph {et~al.}(2006)\citenamefont
  {Nesl\'{a}dek}, \citenamefont {Mare\u{s}}, \citenamefont {Tromson},
  \citenamefont {Mer}, \citenamefont {Bergonzo}, \citenamefont {Hubute\'{i}k},\
  and\ \citenamefont {Kri\u{s}tofik}}]{Nesladek:2006}%
  \BibitemOpen
  \bibfield  {author} {\bibinfo {author} {\bibfnamefont {M.}~\bibnamefont
  {Nesl\'{a}dek}}, \bibinfo {author} {\bibfnamefont {J.}~\bibnamefont
  {Mare\u{s}}}, \bibinfo {author} {\bibfnamefont {D.}~\bibnamefont {Tromson}},
  \bibinfo {author} {\bibfnamefont {C.}~\bibnamefont {Mer}}, \bibinfo {author}
  {\bibfnamefont {P.}~\bibnamefont {Bergonzo}}, \bibinfo {author}
  {\bibfnamefont {P.}~\bibnamefont {Hubute\'{i}k}},\ and\ \bibinfo {author}
  {\bibfnamefont {J.}~\bibnamefont {Kri\u{s}tofik}},\ }\bibfield  {title}
  {\bibinfo {title} {Superconductivity and low temperature electrical transport
  in b-doped cvd nanocrystalline diamond},\ }\href
  {https://doi.org/10.1016/j.stam.2006.04.010} {\bibfield  {journal} {\bibinfo
  {journal} {Sci.\ \& Technol.\ Adv.\ Mat.}\ }\textbf {\bibinfo {volume} {7}},\
  \bibinfo {pages} {S41} (\bibinfo {year} {2006})}\BibitemShut {NoStop}%
\bibitem [{\citenamefont {Sakaguchi}\ \emph {et~al.}(1999)\citenamefont
  {Sakaguchi}, \citenamefont {M.N.-Gamo}, \citenamefont {Kikuchi},
  \citenamefont {Yasu}, \citenamefont {Haneda}, \citenamefont {Suzuki},\ and\
  \citenamefont {Ando}}]{PhysRevB.60.R2139}%
  \BibitemOpen
  \bibfield  {author} {\bibinfo {author} {\bibfnamefont {I.}~\bibnamefont
  {Sakaguchi}}, \bibinfo {author} {\bibnamefont {M.N.-Gamo}}, \bibinfo {author}
  {\bibfnamefont {Y.}~\bibnamefont {Kikuchi}}, \bibinfo {author} {\bibfnamefont
  {E.}~\bibnamefont {Yasu}}, \bibinfo {author} {\bibfnamefont {H.}~\bibnamefont
  {Haneda}}, \bibinfo {author} {\bibfnamefont {T.}~\bibnamefont {Suzuki}},\
  and\ \bibinfo {author} {\bibfnamefont {T.}~\bibnamefont {Ando}},\ }\bibfield
  {title} {\bibinfo {title} {Sulfur: A donor dopant for n-type diamond
  semiconductors},\ }\href {https://doi.org/10.1103/PhysRevB.60.R2139}
  {\bibfield  {journal} {\bibinfo  {journal} {Phys. Rev. B}\ }\textbf {\bibinfo
  {volume} {60}},\ \bibinfo {pages} {R2139} (\bibinfo {year}
  {1999})}\BibitemShut {NoStop}%
\bibitem [{\citenamefont {Koizumi}\ \emph {et~al.}(1998)\citenamefont
  {Koizumi}, \citenamefont {Kamo}, \citenamefont {Sato}, \citenamefont {Mita},
  \citenamefont {Sawabe}, \citenamefont {Reznik}, \citenamefont {Uzan-Saguy},\
  and\ \citenamefont {Kalish}}]{KOIZUMI1998540}%
  \BibitemOpen
  \bibfield  {author} {\bibinfo {author} {\bibfnamefont {S.}~\bibnamefont
  {Koizumi}}, \bibinfo {author} {\bibfnamefont {M.}~\bibnamefont {Kamo}},
  \bibinfo {author} {\bibfnamefont {Y.}~\bibnamefont {Sato}}, \bibinfo {author}
  {\bibfnamefont {S.}~\bibnamefont {Mita}}, \bibinfo {author} {\bibfnamefont
  {A.}~\bibnamefont {Sawabe}}, \bibinfo {author} {\bibfnamefont
  {A.}~\bibnamefont {Reznik}}, \bibinfo {author} {\bibfnamefont
  {C.}~\bibnamefont {Uzan-Saguy}},\ and\ \bibinfo {author} {\bibfnamefont
  {R.}~\bibnamefont {Kalish}},\ }\bibfield  {title} {\bibinfo {title} {Growth
  and characterization of phosphorus doped n-type diamond thin films},\ }\href
  {https://doi.org/https://doi.org/10.1016/S0925-9635(97)00250-1} {\bibfield
  {journal} {\bibinfo  {journal} {Diamond and Related Materials}\ }\textbf
  {\bibinfo {volume} {7}},\ \bibinfo {pages} {540 } (\bibinfo {year}
  {1998})}\BibitemShut {NoStop}%
\bibitem [{\citenamefont {Lin}\ \emph {et~al.}(2011)\citenamefont {Lin},
  \citenamefont {Liao}, \citenamefont {Wei}, \citenamefont {Tsai},
  \citenamefont {Chang},\ and\ \citenamefont {Fang}}]{Lin:2011}%
  \BibitemOpen
  \bibfield  {author} {\bibinfo {author} {\bibfnamefont {C.}~\bibnamefont
  {Lin}}, \bibinfo {author} {\bibfnamefont {W.}~\bibnamefont {Liao}}, \bibinfo
  {author} {\bibfnamefont {D.}~\bibnamefont {Wei}}, \bibinfo {author}
  {\bibfnamefont {J.}~\bibnamefont {Tsai}}, \bibinfo {author} {\bibfnamefont
  {C.}~\bibnamefont {Chang}},\ and\ \bibinfo {author} {\bibfnamefont
  {W.}~\bibnamefont {Fang}},\ }\bibfield  {title} {\bibinfo {title} {Formation
  of ultrananocrystalline diamond films with nitrogen addition},\ }\href
  {https://doi.org/10.1016/j.diamond.2010.12.015} {\bibfield  {journal}
  {\bibinfo  {journal} {Diam.\ \& Rel.\ Mat.}\ }\textbf {\bibinfo {volume}
  {20}},\ \bibinfo {pages} {380} (\bibinfo {year} {2011})}\BibitemShut
  {NoStop}%
\bibitem [{\citenamefont {Bhattacharyya}\ \emph {et~al.}(2001)\citenamefont
  {Bhattacharyya}, \citenamefont {Auciello}, \citenamefont {Birrell},
  \citenamefont {Carlisle}, \citenamefont {Curtiss}, \citenamefont {Goyette},
  \citenamefont {Gruen}, \citenamefont {Krauss}, \citenamefont {Schlueter},
  \citenamefont {Sumant},\ and\ \citenamefont {Zapol}}]{Bhattacharyya:2001}%
  \BibitemOpen
  \bibfield  {author} {\bibinfo {author} {\bibfnamefont {S.}~\bibnamefont
  {Bhattacharyya}}, \bibinfo {author} {\bibfnamefont {O.}~\bibnamefont
  {Auciello}}, \bibinfo {author} {\bibfnamefont {J.}~\bibnamefont {Birrell}},
  \bibinfo {author} {\bibfnamefont {J.}~\bibnamefont {Carlisle}}, \bibinfo
  {author} {\bibfnamefont {L.}~\bibnamefont {Curtiss}}, \bibinfo {author}
  {\bibfnamefont {A.}~\bibnamefont {Goyette}}, \bibinfo {author} {\bibfnamefont
  {D.}~\bibnamefont {Gruen}}, \bibinfo {author} {\bibfnamefont
  {A.}~\bibnamefont {Krauss}}, \bibinfo {author} {\bibfnamefont
  {J.}~\bibnamefont {Schlueter}}, \bibinfo {author} {\bibfnamefont
  {A.}~\bibnamefont {Sumant}},\ and\ \bibinfo {author} {\bibfnamefont
  {P.}~\bibnamefont {Zapol}},\ }\bibfield  {title} {\bibinfo {title} {Synthesis
  and characterization of highly-conducting nitrogen-doped ultrananocrystalline
  diamond films},\ }\href {https://doi.org/10.1063/1.1400761} {\bibfield
  {journal} {\bibinfo  {journal} {Appl.\ Phys.\ Lett.}\ }\textbf {\bibinfo
  {volume} {79}},\ \bibinfo {pages} {1441} (\bibinfo {year}
  {2001})}\BibitemShut {NoStop}%
\bibitem [{\citenamefont {Birrell}\ \emph {et~al.}(2002)\citenamefont
  {Birrell}, \citenamefont {Carlisle}, \citenamefont {Auciello}, \citenamefont
  {Gruen},\ and\ \citenamefont {Gibson}}]{Birell:2002}%
  \BibitemOpen
  \bibfield  {author} {\bibinfo {author} {\bibfnamefont {J.}~\bibnamefont
  {Birrell}}, \bibinfo {author} {\bibfnamefont {J.~A.}\ \bibnamefont
  {Carlisle}}, \bibinfo {author} {\bibfnamefont {O.}~\bibnamefont {Auciello}},
  \bibinfo {author} {\bibfnamefont {D.}~\bibnamefont {Gruen}},\ and\ \bibinfo
  {author} {\bibfnamefont {J.}~\bibnamefont {Gibson}},\ }\bibfield  {title}
  {\bibinfo {title} {Morphology and electronic structure in nitrogen-doped
  ultrananocrystalline diamond},\ }\href {https://doi.org/10.1063/1.1503153}
  {\bibfield  {journal} {\bibinfo  {journal} {Appl.\ Phys.\ Lett.}\ }\textbf
  {\bibinfo {volume} {81}},\ \bibinfo {pages} {2235} (\bibinfo {year}
  {2002})}\BibitemShut {NoStop}%
\bibitem [{\citenamefont {Zapol}\ \emph {et~al.}(2001)\citenamefont {Zapol},
  \citenamefont {Sternberg}, \citenamefont {Curtiss}, \citenamefont
  {Frauenheim},\ and\ \citenamefont {Gruen}}]{Zapol:2001}%
  \BibitemOpen
  \bibfield  {author} {\bibinfo {author} {\bibfnamefont {P.}~\bibnamefont
  {Zapol}}, \bibinfo {author} {\bibfnamefont {M.}~\bibnamefont {Sternberg}},
  \bibinfo {author} {\bibfnamefont {L.}~\bibnamefont {Curtiss}}, \bibinfo
  {author} {\bibfnamefont {T.}~\bibnamefont {Frauenheim}},\ and\ \bibinfo
  {author} {\bibfnamefont {D.}~\bibnamefont {Gruen}},\ }\bibfield  {title}
  {\bibinfo {title} {Tight-binding molecular-dynamics simulation of impurities
  in ultrananocrystalline diamond grain boundaries},\ }\href
  {https://doi.org/10.1103/PhysRevB.65.045403} {\bibfield  {journal} {\bibinfo
  {journal} {Phys.\ Rev.\ B}\ }\textbf {\bibinfo {volume} {65}},\ \bibinfo
  {pages} {045403} (\bibinfo {year} {2001})}\BibitemShut {NoStop}%
\bibitem [{\citenamefont {Beloborodov}\ \emph {et~al.}(2006)\citenamefont
  {Beloborodov}, \citenamefont {Zapol}, \citenamefont {Gruen},\ and\
  \citenamefont {Curtiss}}]{Beloborodov:2006}%
  \BibitemOpen
  \bibfield  {author} {\bibinfo {author} {\bibfnamefont {I.}~\bibnamefont
  {Beloborodov}}, \bibinfo {author} {\bibfnamefont {P.}~\bibnamefont {Zapol}},
  \bibinfo {author} {\bibfnamefont {D.}~\bibnamefont {Gruen}},\ and\ \bibinfo
  {author} {\bibfnamefont {L.}~\bibnamefont {Curtiss}},\ }\bibfield  {title}
  {\bibinfo {title} {Transport properties of n-type ultrananocrystalline
  diamond films},\ }\href {https://doi.org/10.1103/PhysRevB.74.235434}
  {\bibfield  {journal} {\bibinfo  {journal} {Phys.\ Rev.\ B}\ }\textbf
  {\bibinfo {volume} {74}},\ \bibinfo {pages} {235434} (\bibinfo {year}
  {2006})}\BibitemShut {NoStop}%
\bibitem [{\citenamefont {Bhattacharyya}(2008)}]{Bhattacharyya:2008}%
  \BibitemOpen
  \bibfield  {author} {\bibinfo {author} {\bibfnamefont {S.}~\bibnamefont
  {Bhattacharyya}},\ }\bibfield  {title} {\bibinfo {title} {Two-dimensional
  transport in disordered carbon and nanocrystalline diamond films},\ }\href
  {https://doi.org/10.1103/PhysRevB.77.233407} {\bibfield  {journal} {\bibinfo
  {journal} {Phys.\ Rev.\ B}\ }\textbf {\bibinfo {volume} {77}},\ \bibinfo
  {pages} {233407} (\bibinfo {year} {2008})}\BibitemShut {NoStop}%
\bibitem [{\citenamefont {Shah}\ \emph {et~al.}(2010)\citenamefont {Shah},
  \citenamefont {Churochkin}, \citenamefont {Chiguvare},\ and\ \citenamefont
  {Bhattacharyya}}]{Shah:2010}%
  \BibitemOpen
  \bibfield  {author} {\bibinfo {author} {\bibfnamefont {K.}~\bibnamefont
  {Shah}}, \bibinfo {author} {\bibfnamefont {D.}~\bibnamefont {Churochkin}},
  \bibinfo {author} {\bibfnamefont {Z.}~\bibnamefont {Chiguvare}},\ and\
  \bibinfo {author} {\bibfnamefont {S.}~\bibnamefont {Bhattacharyya}},\
  }\bibfield  {title} {\bibinfo {title} {Anisotropic weakly localized transport
  in nitrogen-doped ultrananocrystalline diamond films},\ }\href
  {https://doi.org/10.1103/PhysRevB.82.184206} {\bibfield  {journal} {\bibinfo
  {journal} {Phys.\ Rev.\ B}\ }\textbf {\bibinfo {volume} {82}},\ \bibinfo
  {pages} {184206} (\bibinfo {year} {2010})}\BibitemShut {NoStop}%
\bibitem [{\citenamefont {Churochkin}\ and\ \citenamefont
  {Bhattacharyya}(2012)}]{Churouchkin:2012}%
  \BibitemOpen
  \bibfield  {author} {\bibinfo {author} {\bibfnamefont {D.}~\bibnamefont
  {Churochkin}}\ and\ \bibinfo {author} {\bibfnamefont {S.}~\bibnamefont
  {Bhattacharyya}},\ }\bibfield  {title} {\bibinfo {title} {Tuneable
  anisotropic transport in nitrogen-doped nanocrystalline diamond films:
  Evidence of a graphite-diamond hybrid superlattice},\ }\href
  {https://doi.org/10.1209/0295-5075/100/67004} {\bibfield  {journal} {\bibinfo
   {journal} {EPL}\ }\textbf {\bibinfo {volume} {100}},\ \bibinfo {pages}
  {67004} (\bibinfo {year} {2012})}\BibitemShut {NoStop}%
\bibitem [{Note1()}]{Note1}%
  \BibitemOpen
  \bibinfo {note} {Advanced Diamond Technologies Inc., Romeoville, IL, USA.
  (www.thindiamond.com)}\BibitemShut {NoStop}%
\bibitem [{\citenamefont {Arenal}\ \emph {et~al.}(2007)\citenamefont {Arenal},
  \citenamefont {Bruno}, \citenamefont {Miller}, \citenamefont {Bleuel},
  \citenamefont {Lal},\ and\ \citenamefont {Gruen}}]{Arenal:2007}%
  \BibitemOpen
  \bibfield  {author} {\bibinfo {author} {\bibfnamefont {R.}~\bibnamefont
  {Arenal}}, \bibinfo {author} {\bibfnamefont {P.}~\bibnamefont {Bruno}},
  \bibinfo {author} {\bibfnamefont {D.}~\bibnamefont {Miller}}, \bibinfo
  {author} {\bibfnamefont {M.}~\bibnamefont {Bleuel}}, \bibinfo {author}
  {\bibfnamefont {J.}~\bibnamefont {Lal}},\ and\ \bibinfo {author}
  {\bibfnamefont {D.}~\bibnamefont {Gruen}},\ }\bibfield  {title} {\bibinfo
  {title} {Diamond nanowires and the insulator-metal transition in
  ultrananocrystalline diamond films},\ }\href
  {https://doi.org/10.1103/PhysRevB.75.195431} {\bibfield  {journal} {\bibinfo
  {journal} {Phys.\ Rev.\ B}\ }\textbf {\bibinfo {volume} {75}},\ \bibinfo
  {pages} {195431} (\bibinfo {year} {2007})}\BibitemShut {NoStop}%
\bibitem [{\citenamefont {Vlasov}\ \emph {et~al.}(2007)\citenamefont {Vlasov},
  \citenamefont {Lebedev}, \citenamefont {Ralchenko}, \citenamefont
  {Goovaerts}, \citenamefont {Bertoni}, \citenamefont {van Tendeloo},\ and\
  \citenamefont {Konov}}]{Vlasov:2007}%
  \BibitemOpen
  \bibfield  {author} {\bibinfo {author} {\bibfnamefont {I.}~\bibnamefont
  {Vlasov}}, \bibinfo {author} {\bibfnamefont {O.}~\bibnamefont {Lebedev}},
  \bibinfo {author} {\bibfnamefont {V.}~\bibnamefont {Ralchenko}}, \bibinfo
  {author} {\bibfnamefont {E.}~\bibnamefont {Goovaerts}}, \bibinfo {author}
  {\bibfnamefont {G.}~\bibnamefont {Bertoni}}, \bibinfo {author} {\bibfnamefont
  {G.}~\bibnamefont {van Tendeloo}},\ and\ \bibinfo {author} {\bibfnamefont
  {V.}~\bibnamefont {Konov}},\ }\bibfield  {title} {\bibinfo {title} {Hybrid
  diamond-graphite nanowires produced by microwave plasma chemical vapor
  deposition},\ }\href {https://doi.org/10.1002/adma.200700442} {\bibfield
  {journal} {\bibinfo  {journal} {Adv.\ Mat.}\ }\textbf {\bibinfo {volume}
  {19}},\ \bibinfo {pages} {4058} (\bibinfo {year} {2007})}\BibitemShut
  {NoStop}%
\bibitem [{\citenamefont {Tzeng}\ \emph {et~al.}(2014)\citenamefont {Tzeng},
  \citenamefont {Yeh}, \citenamefont {Fang},\ and\ \citenamefont
  {Chu}}]{Tzeng:2014}%
  \BibitemOpen
  \bibfield  {author} {\bibinfo {author} {\bibfnamefont {Y.}~\bibnamefont
  {Tzeng}}, \bibinfo {author} {\bibfnamefont {S.}~\bibnamefont {Yeh}}, \bibinfo
  {author} {\bibfnamefont {W.}~\bibnamefont {Fang}},\ and\ \bibinfo {author}
  {\bibfnamefont {Y.}~\bibnamefont {Chu}},\ }\bibfield  {title} {\bibinfo
  {title} {Nitrogen-incorporated ultrananocrystalline diamond and
  multi-layer-graphene-like hybrid carbon films},\ }\href
  {https://doi.org/10.1038/srep04531} {\bibfield  {journal} {\bibinfo
  {journal} {Scientific Reports}\ }\textbf {\bibinfo {volume} {4}},\ \bibinfo
  {pages} {4531} (\bibinfo {year} {2014})}\BibitemShut {NoStop}%
\bibitem [{\citenamefont {Birrell}\ \emph {et~al.}(2003)\citenamefont
  {Birrell}, \citenamefont {Gerbi}, \citenamefont {Auciello}, \citenamefont
  {Gibson}, \citenamefont {Gruen},\ and\ \citenamefont
  {Carlisle}}]{BirrellBondingStructure}%
  \BibitemOpen
  \bibfield  {author} {\bibinfo {author} {\bibfnamefont {J.}~\bibnamefont
  {Birrell}}, \bibinfo {author} {\bibfnamefont {J.~E.}\ \bibnamefont {Gerbi}},
  \bibinfo {author} {\bibfnamefont {O.}~\bibnamefont {Auciello}}, \bibinfo
  {author} {\bibfnamefont {J.~M.}\ \bibnamefont {Gibson}}, \bibinfo {author}
  {\bibfnamefont {D.~M.}\ \bibnamefont {Gruen}},\ and\ \bibinfo {author}
  {\bibfnamefont {J.~A.}\ \bibnamefont {Carlisle}},\ }\bibfield  {title}
  {\bibinfo {title} {Bonding structure in nitrogen doped ultrananocrystalline
  diamond},\ }\href {https://doi.org/10.1063/1.1564880} {\bibfield  {journal}
  {\bibinfo  {journal} {Journal of Applied Physics}\ }\textbf {\bibinfo
  {volume} {93}},\ \bibinfo {pages} {5606} (\bibinfo {year}
  {2003})}\BibitemShut {NoStop}%
\bibitem [{\citenamefont {Zolotukhin}\ \emph {et~al.}(2013)\citenamefont
  {Zolotukhin}, \citenamefont {Dolganov},\ and\ \citenamefont
  {Obraztsov}}]{ZOLOTUKHIN201364}%
  \BibitemOpen
  \bibfield  {author} {\bibinfo {author} {\bibfnamefont {A.~A.}\ \bibnamefont
  {Zolotukhin}}, \bibinfo {author} {\bibfnamefont {M.~A.}\ \bibnamefont
  {Dolganov}},\ and\ \bibinfo {author} {\bibfnamefont {A.~N.}\ \bibnamefont
  {Obraztsov}},\ }\bibfield  {title} {\bibinfo {title} {Nanodiamond films with
  dendrite structure formed by needle crystallites},\ }\href
  {https://doi.org/https://doi.org/10.1016/j.diamond.2013.04.003} {\bibfield
  {journal} {\bibinfo  {journal} {Diamond and Related Materials}\ }\textbf
  {\bibinfo {volume} {37}},\ \bibinfo {pages} {64 } (\bibinfo {year}
  {2013})}\BibitemShut {NoStop}%
\bibitem [{\citenamefont {Gruen}(1999)}]{annurev.matsci.29.1.211}%
  \BibitemOpen
  \bibfield  {author} {\bibinfo {author} {\bibfnamefont {D.~M.}\ \bibnamefont
  {Gruen}},\ }\bibfield  {title} {\bibinfo {title} {Nanocrystalline diamond
  films},\ }\href {https://doi.org/10.1146/annurev.matsci.29.1.211} {\bibfield
  {journal} {\bibinfo  {journal} {Annual Review of Materials Science}\ }\textbf
  {\bibinfo {volume} {29}},\ \bibinfo {pages} {211} (\bibinfo {year}
  {1999})}\BibitemShut {NoStop}%
\bibitem [{\citenamefont {Eikenberg}(2015)}]{Ninathesis}%
  \BibitemOpen
  \bibfield  {author} {\bibinfo {author} {\bibfnamefont {N.}~\bibnamefont
  {Eikenberg}},\ }\emph {\bibinfo {title} {Characterization of silicon and
  diamond semiconductor devices in the low temperature regime}},\ \href@noop {}
  {Ph.D. thesis},\ \bibinfo  {school} {School of Physics, University of
  Melbourne} (\bibinfo {year} {2015})\BibitemShut {NoStop}%
\bibitem [{\citenamefont {Garratt}(2013)}]{Garratt:2013}%
  \BibitemOpen
  \bibfield  {author} {\bibinfo {author} {\bibfnamefont {E.}~\bibnamefont
  {Garratt}},\ }\emph {\bibinfo {title} {Compositional and structural analysis
  of nitrogen incorporated and ion implanted diamond thin films}},\ \href@noop
  {} {Ph.D. thesis},\ \bibinfo  {school} {Department of Physics, Western
  Michigan University} (\bibinfo {year} {2013})\BibitemShut {NoStop}%
\bibitem [{\citenamefont {Williams}\ \emph {et~al.}(2004)\citenamefont
  {Williams}, \citenamefont {Curat}, \citenamefont {Gerbi}, \citenamefont
  {Gruen},\ and\ \citenamefont {Jackman}}]{Williams:2004}%
  \BibitemOpen
  \bibfield  {author} {\bibinfo {author} {\bibfnamefont {O.}~\bibnamefont
  {Williams}}, \bibinfo {author} {\bibfnamefont {S.}~\bibnamefont {Curat}},
  \bibinfo {author} {\bibfnamefont {J.}~\bibnamefont {Gerbi}}, \bibinfo
  {author} {\bibfnamefont {D.}~\bibnamefont {Gruen}},\ and\ \bibinfo {author}
  {\bibfnamefont {R.}~\bibnamefont {Jackman}},\ }\bibfield  {title} {\bibinfo
  {title} {$n$-type conductivity in ultrananocrystalline diamond films},\
  }\href {https://doi.org/10.1063/1.1785288} {\bibfield  {journal} {\bibinfo
  {journal} {Appl.\ Phys.\ Lett.}\ }\textbf {\bibinfo {volume} {85}},\ \bibinfo
  {pages} {1680} (\bibinfo {year} {2004})}\BibitemShut {NoStop}%
\bibitem [{\citenamefont {Mare$\breve{s}$}\ \emph {et~al.}(2006)\citenamefont
  {Mare$\breve{s}$}, \citenamefont {Hub$\acute{i}$k}, \citenamefont
  {J.~Kri$\breve{s}$tofik}, \citenamefont {Fanta}, \citenamefont
  {Nesl$\acute{a}$dek}, \citenamefont {Williams},\ and\ \citenamefont
  {Gruen}}]{Mares:2006}%
  \BibitemOpen
  \bibfield  {author} {\bibinfo {author} {\bibfnamefont {J.}~\bibnamefont
  {Mare$\breve{s}$}}, \bibinfo {author} {\bibfnamefont {P.}~\bibnamefont
  {Hub$\acute{i}$k}}, \bibinfo {author} {\bibfnamefont {D.~K.}\ \bibnamefont
  {J.~Kri$\breve{s}$tofik}}, \bibinfo {author} {\bibfnamefont {M.}~\bibnamefont
  {Fanta}}, \bibinfo {author} {\bibfnamefont {M.}~\bibnamefont
  {Nesl$\acute{a}$dek}}, \bibinfo {author} {\bibfnamefont {O.}~\bibnamefont
  {Williams}},\ and\ \bibinfo {author} {\bibfnamefont {D.}~\bibnamefont
  {Gruen}},\ }\bibfield  {title} {\bibinfo {title} {Weak localization in
  ultrananocrystalline diamond},\ }\href {https://doi.org/10.1063/1.2176853}
  {\bibfield  {journal} {\bibinfo  {journal} {Appl.\ Phys.\ Lett.}\ }\textbf
  {\bibinfo {volume} {88}},\ \bibinfo {pages} {092107} (\bibinfo {year}
  {2006})}\BibitemShut {NoStop}%
\bibitem [{\citenamefont {Han}\ \emph {et~al.}(2010)\citenamefont {Han},
  \citenamefont {Brant},\ and\ \citenamefont {Kim}}]{Han:2010aa}%
  \BibitemOpen
  \bibfield  {author} {\bibinfo {author} {\bibfnamefont {M.~Y.}\ \bibnamefont
  {Han}}, \bibinfo {author} {\bibfnamefont {J.~C.}\ \bibnamefont {Brant}},\
  and\ \bibinfo {author} {\bibfnamefont {P.}~\bibnamefont {Kim}},\ }\bibfield
  {title} {\bibinfo {title} {Electron transport in disordered graphene
  nanoribbons},\ }\bibfield  {journal} {\bibinfo  {journal} {Phys. Rev. Lett.}\
  }\textbf {\bibinfo {volume} {104}},\ \href
  {https://doi.org/10.1103/PhysRevLett.104.056801}
  {10.1103/PhysRevLett.104.056801} (\bibinfo {year} {2010})\BibitemShut
  {NoStop}%
\bibitem [{\citenamefont {Kim}\ \emph {et~al.}(2016)\citenamefont {Kim},
  \citenamefont {Lara-Avila}, \citenamefont {Kang}, \citenamefont {He},
  \citenamefont {Eklof}, \citenamefont {Hong}, \citenamefont {Park},
  \citenamefont {Moth-Poulsen}, \citenamefont {Matsushita}, \citenamefont
  {Akagi}, \citenamefont {Kubatkin},\ and\ \citenamefont {Park}}]{Kim:2016aa}%
  \BibitemOpen
  \bibfield  {author} {\bibinfo {author} {\bibfnamefont {K.~H.}\ \bibnamefont
  {Kim}}, \bibinfo {author} {\bibfnamefont {S.}~\bibnamefont {Lara-Avila}},
  \bibinfo {author} {\bibfnamefont {H.}~\bibnamefont {Kang}}, \bibinfo {author}
  {\bibfnamefont {H.}~\bibnamefont {He}}, \bibinfo {author} {\bibfnamefont
  {J.}~\bibnamefont {Eklof}}, \bibinfo {author} {\bibfnamefont {S.~J.}\
  \bibnamefont {Hong}}, \bibinfo {author} {\bibfnamefont {M.}~\bibnamefont
  {Park}}, \bibinfo {author} {\bibfnamefont {K.}~\bibnamefont {Moth-Poulsen}},
  \bibinfo {author} {\bibfnamefont {S.}~\bibnamefont {Matsushita}}, \bibinfo
  {author} {\bibfnamefont {K.}~\bibnamefont {Akagi}}, \bibinfo {author}
  {\bibfnamefont {S.}~\bibnamefont {Kubatkin}},\ and\ \bibinfo {author}
  {\bibfnamefont {Y.~W.}\ \bibnamefont {Park}},\ }\bibfield  {title} {\bibinfo
  {title} {Apparent power law scaling of variable range hopping conduction in
  carbonized polymer nanofibers},\ }\bibfield  {journal} {\bibinfo  {journal}
  {Sci Rep}\ }\textbf {\bibinfo {volume} {6}},\ \href
  {https://doi.org/10.1038/srep37783} {10.1038/srep37783} (\bibinfo {year}
  {2016})\BibitemShut {NoStop}%
\bibitem [{\citenamefont {Yu}\ \emph {et~al.}(2004)\citenamefont {Yu},
  \citenamefont {Wang}, \citenamefont {Wehrenberg},\ and\ \citenamefont
  {Guyot-Sionnest}}]{Yu:2004aa}%
  \BibitemOpen
  \bibfield  {author} {\bibinfo {author} {\bibfnamefont {D.}~\bibnamefont
  {Yu}}, \bibinfo {author} {\bibfnamefont {C.~J.}\ \bibnamefont {Wang}},
  \bibinfo {author} {\bibfnamefont {B.~L.}\ \bibnamefont {Wehrenberg}},\ and\
  \bibinfo {author} {\bibfnamefont {P.}~\bibnamefont {Guyot-Sionnest}},\
  }\bibfield  {title} {\bibinfo {title} {Variable range hopping conduction in
  semiconductor nanocrystal solids},\ }\bibfield  {journal} {\bibinfo
  {journal} {Phys. Rev. Lett.}\ }\textbf {\bibinfo {volume} {92}},\ \href
  {https://doi.org/10.1103/PhysRevLett.92.216802}
  {10.1103/PhysRevLett.92.216802} (\bibinfo {year} {2004})\BibitemShut
  {NoStop}%
\bibitem [{\citenamefont {Grabert}\ and\ \citenamefont
  {Devoret}(1992)}]{GrabertDevoret:1992}%
  \BibitemOpen
  \bibfield  {author} {\bibinfo {author} {\bibfnamefont {H.}~\bibnamefont
  {Grabert}}\ and\ \bibinfo {author} {\bibfnamefont {M.}~\bibnamefont
  {Devoret}},\ }\href@noop {} {\emph {\bibinfo {title} {Single charge
  tunneling: Coulomb blockade phenomena in nanostructures}}}\ (\bibinfo
  {publisher} {Springer US},\ \bibinfo {year} {1992})\BibitemShut {NoStop}%
\bibitem [{\citenamefont {Joung}\ and\ \citenamefont
  {Khondaker}(2013)}]{Joung:2013aa}%
  \BibitemOpen
  \bibfield  {author} {\bibinfo {author} {\bibfnamefont {D.}~\bibnamefont
  {Joung}}\ and\ \bibinfo {author} {\bibfnamefont {S.~I.}\ \bibnamefont
  {Khondaker}},\ }\bibfield  {title} {\bibinfo {title} {Structural evolution of
  reduced graphene oxide of varying carbon sp2 fractions investigated via
  coulomb blockade transport},\ }\href@noop {} {\bibfield  {journal} {\bibinfo
  {journal} {The Journal of Physical Chemistry C}\ }\textbf {\bibinfo {volume}
  {117}},\ \bibinfo {pages} {26776} (\bibinfo {year} {2013})}\BibitemShut
  {NoStop}%
\bibitem [{\citenamefont {Romero}\ and\ \citenamefont
  {Drndic}(2005)}]{PhysRevLett.95.156801}%
  \BibitemOpen
  \bibfield  {author} {\bibinfo {author} {\bibfnamefont {H.~E.}\ \bibnamefont
  {Romero}}\ and\ \bibinfo {author} {\bibfnamefont {M.}~\bibnamefont
  {Drndic}},\ }\bibfield  {title} {\bibinfo {title} {Coulomb blockade and
  hopping conduction in pbse quantum dots},\ }\href
  {https://doi.org/10.1103/PhysRevLett.95.156801} {\bibfield  {journal}
  {\bibinfo  {journal} {Phys. Rev. Lett.}\ }\textbf {\bibinfo {volume} {95}},\
  \bibinfo {pages} {156801} (\bibinfo {year} {2005})}\BibitemShut {NoStop}%
\bibitem [{\citenamefont {Black}\ \emph {et~al.}(2000)\citenamefont {Black},
  \citenamefont {Murray}, \citenamefont {Sandstrom},\ and\ \citenamefont
  {Sun}}]{Black1131}%
  \BibitemOpen
  \bibfield  {author} {\bibinfo {author} {\bibfnamefont {C.~T.}\ \bibnamefont
  {Black}}, \bibinfo {author} {\bibfnamefont {C.~B.}\ \bibnamefont {Murray}},
  \bibinfo {author} {\bibfnamefont {R.~L.}\ \bibnamefont {Sandstrom}},\ and\
  \bibinfo {author} {\bibfnamefont {S.}~\bibnamefont {Sun}},\ }\bibfield
  {title} {\bibinfo {title} {Spin-dependent tunneling in self-assembled
  cobalt-nanocrystal superlattices},\ }\href
  {https://doi.org/10.1126/science.290.5494.1131} {\bibfield  {journal}
  {\bibinfo  {journal} {Science}\ }\textbf {\bibinfo {volume} {290}},\ \bibinfo
  {pages} {1131} (\bibinfo {year} {2000})}\BibitemShut {NoStop}%
\bibitem [{\citenamefont {Auciello}\ and\ \citenamefont
  {Sumant}(2010)}]{AUCIELLO2010699}%
  \BibitemOpen
  \bibfield  {author} {\bibinfo {author} {\bibfnamefont {O.}~\bibnamefont
  {Auciello}}\ and\ \bibinfo {author} {\bibfnamefont {A.~V.}\ \bibnamefont
  {Sumant}},\ }\bibfield  {title} {\bibinfo {title} {Status review of the
  science and technology of ultrananocrystalline diamond (uncd{\texttrademark})
  films and application to multifunctional devices},\ }\href
  {https://doi.org/https://doi.org/10.1016/j.diamond.2010.03.015} {\bibfield
  {journal} {\bibinfo  {journal} {Diamond and Related Materials}\ }\textbf
  {\bibinfo {volume} {19}},\ \bibinfo {pages} {699 } (\bibinfo {year}
  {2010})}\BibitemShut {NoStop}%
\bibitem [{\citenamefont {Nesl$\acute{a}$dek}\ \emph
  {et~al.}(2006)\citenamefont {Nesl$\acute{a}$dek}, \citenamefont {Tromson},
  \citenamefont {Bergonzo}, \citenamefont {Hub$\acute{i}$k}, \citenamefont
  {Mare$\breve{s}$}, \citenamefont {Kri$\breve{s}$tofik}, \citenamefont
  {Kindl}, \citenamefont {Williams},\ and\ \citenamefont
  {Gruen}}]{Nesladek:2006b}%
  \BibitemOpen
  \bibfield  {author} {\bibinfo {author} {\bibfnamefont {M.}~\bibnamefont
  {Nesl$\acute{a}$dek}}, \bibinfo {author} {\bibfnamefont {D.}~\bibnamefont
  {Tromson}}, \bibinfo {author} {\bibfnamefont {P.}~\bibnamefont {Bergonzo}},
  \bibinfo {author} {\bibfnamefont {P.}~\bibnamefont {Hub$\acute{i}$k}},
  \bibinfo {author} {\bibfnamefont {J.}~\bibnamefont {Mare$\breve{s}$}},
  \bibinfo {author} {\bibfnamefont {J.}~\bibnamefont {Kri$\breve{s}$tofik}},
  \bibinfo {author} {\bibfnamefont {D.}~\bibnamefont {Kindl}}, \bibinfo
  {author} {\bibfnamefont {O.}~\bibnamefont {Williams}},\ and\ \bibinfo
  {author} {\bibfnamefont {D.}~\bibnamefont {Gruen}},\ }\bibfield  {title}
  {\bibinfo {title} {Low-temperature magnetoresistance study of electrical
  transport in n- and b-doped ultrananocrystalline and nanocrystalline diamond
  films},\ }\href {https://doi.org/10.1016/j.diamond.2005.11.001} {\bibfield
  {journal} {\bibinfo  {journal} {Diam.\ \& Rel.\ Mat.}\ }\textbf {\bibinfo
  {volume} {15}},\ \bibinfo {pages} {607} (\bibinfo {year} {2006})}\BibitemShut
  {NoStop}%
\bibitem [{\citenamefont {{Willems van Beveren}}\ \emph
  {et~al.}(2016)\citenamefont {{Willems van Beveren}}, \citenamefont {Liu},
  \citenamefont {Bowers}, \citenamefont {Ganesan}, \citenamefont {Johnson},
  \citenamefont {McCallum},\ and\ \citenamefont
  {Prawer}}]{Willemsvanbeveren:2016}%
  \BibitemOpen
  \bibfield  {author} {\bibinfo {author} {\bibfnamefont {L.}~\bibnamefont
  {{Willems van Beveren}}}, \bibinfo {author} {\bibfnamefont {R.}~\bibnamefont
  {Liu}}, \bibinfo {author} {\bibfnamefont {H.}~\bibnamefont {Bowers}},
  \bibinfo {author} {\bibfnamefont {K.}~\bibnamefont {Ganesan}}, \bibinfo
  {author} {\bibfnamefont {B.}~\bibnamefont {Johnson}}, \bibinfo {author}
  {\bibfnamefont {J.}~\bibnamefont {McCallum}},\ and\ \bibinfo {author}
  {\bibfnamefont {S.}~\bibnamefont {Prawer}},\ }\bibfield  {title} {\bibinfo
  {title} {Optical and electronic properties of sub-surface conducting layers
  in diamond created by mev b-implantation at elevated temperatures},\ }\href
  {https://doi.org/10.1063/1.4953583} {\bibfield  {journal} {\bibinfo
  {journal} {J. Appl.\ Phys.}\ }\textbf {\bibinfo {volume} {119}},\ \bibinfo
  {pages} {223902} (\bibinfo {year} {2016})}\BibitemShut {NoStop}%
\bibitem [{\citenamefont {Abeles}\ \emph {et~al.}(1975)\citenamefont {Abeles},
  \citenamefont {Sheng}, \citenamefont {Coutts},\ and\ \citenamefont
  {Arie}}]{Abeles:1975}%
  \BibitemOpen
  \bibfield  {author} {\bibinfo {author} {\bibfnamefont {B.}~\bibnamefont
  {Abeles}}, \bibinfo {author} {\bibfnamefont {P.}~\bibnamefont {Sheng}},
  \bibinfo {author} {\bibfnamefont {M.}~\bibnamefont {Coutts}},\ and\ \bibinfo
  {author} {\bibfnamefont {Y.}~\bibnamefont {Arie}},\ }\bibfield  {title}
  {\bibinfo {title} {Structural and electrical properties of granular metal
  films},\ }\href {https://doi.org/10.1080/00018737500101431} {\bibfield
  {journal} {\bibinfo  {journal} {Adv.\ Phys.}\ }\textbf {\bibinfo {volume}
  {24}},\ \bibinfo {pages} {407} (\bibinfo {year} {1975})}\BibitemShut
  {NoStop}%
\bibitem [{\citenamefont {Shklovskii}\ and\ \citenamefont
  {Efros}(1984)}]{ShklovskiiEfros:1984}%
  \BibitemOpen
  \bibfield  {author} {\bibinfo {author} {\bibfnamefont {B.}~\bibnamefont
  {Shklovskii}}\ and\ \bibinfo {author} {\bibfnamefont {A.}~\bibnamefont
  {Efros}},\ }\href@noop {} {\emph {\bibinfo {title} {Electronic properties of
  doped semiconductors}}}\ (\bibinfo  {publisher} {Springer Verlag, Berlin},\
  \bibinfo {year} {1984})\BibitemShut {NoStop}%
\bibitem [{\citenamefont {Choy}\ \emph {et~al.}(2008)\citenamefont {Choy},
  \citenamefont {Stoneham}, \citenamefont {Ortuno},\ and\ \citenamefont
  {Somoza}}]{Choy:2008}%
  \BibitemOpen
  \bibfield  {author} {\bibinfo {author} {\bibfnamefont {T.}~\bibnamefont
  {Choy}}, \bibinfo {author} {\bibfnamefont {A.}~\bibnamefont {Stoneham}},
  \bibinfo {author} {\bibfnamefont {M.}~\bibnamefont {Ortuno}},\ and\ \bibinfo
  {author} {\bibfnamefont {A.}~\bibnamefont {Somoza}},\ }\bibfield  {title}
  {\bibinfo {title} {Negative magnetoresistance in ultrananocrystalline
  diamond: strong or weak localizaiton?},\ }\href
  {https://doi.org/10.1063/1.2826542} {\bibfield  {journal} {\bibinfo
  {journal} {Appl.\ Phys.\ Lett.}\ }\textbf {\bibinfo {volume} {92}},\ \bibinfo
  {pages} {012120} (\bibinfo {year} {2008})}\BibitemShut {NoStop}%
\bibitem [{\citenamefont {Groves}\ and\ \citenamefont
  {Martin}(1940)}]{TF9403500575}%
  \BibitemOpen
  \bibfield  {author} {\bibinfo {author} {\bibfnamefont {L.}~\bibnamefont
  {Groves}}\ and\ \bibinfo {author} {\bibfnamefont {A.}~\bibnamefont
  {Martin}},\ }\bibfield  {title} {\bibinfo {title} {The dielectric constant of
  diamond},\ }\href {https://doi.org/10.1039/TF9403500575} {\bibfield
  {journal} {\bibinfo  {journal} {Trans. Faraday Soc.}\ }\textbf {\bibinfo
  {volume} {35}},\ \bibinfo {pages} {575} (\bibinfo {year} {1940})}\BibitemShut
  {NoStop}%
\bibitem [{\citenamefont {Govindaraju}\ and\ \citenamefont
  {Singh}(2015)}]{epsilonUNCD}%
  \BibitemOpen
  \bibfield  {author} {\bibinfo {author} {\bibfnamefont {N.}~\bibnamefont
  {Govindaraju}}\ and\ \bibinfo {author} {\bibfnamefont {R.}~\bibnamefont
  {Singh}},\ }\bibinfo {title} {Dielectric properties and applications of
  nanocrystalline diamond thin films},\ in\ \href
  {https://doi.org/10.1002/9781119183860.ch15} {\emph {\bibinfo {booktitle}
  {Processing and Properties of Advanced Ceramics and Composites VII}}},\
  \bibinfo {editor} {edited by\ \bibinfo {editor} {\bibfnamefont {N.~P. B. J.
  P. S. R. H. R. C. N. J. M. G. P. S. J. G. B. G.~S.}\ \bibnamefont {Morsi
  M.~Mahmoud}, \bibfnamefont {Amar~Bhalla}}\ and\ \bibinfo {editor}
  {\bibfnamefont {D.}~\bibnamefont {Zhu}}}\ (\bibinfo  {publisher} {John Wiley
  \& Sons Ltd.},\ \bibinfo {year} {2015})\ Chap.~\bibinfo {chapter} {15}, pp.\
  \bibinfo {pages} {137--150}\BibitemShut {NoStop}%
\bibitem [{\citenamefont {Entin-Wohlman}\ \emph {et~al.}(1983)\citenamefont
  {Entin-Wohlman}, \citenamefont {Gefen},\ and\ \citenamefont
  {Shapira}}]{Entin_Wohlman_1983}%
  \BibitemOpen
  \bibfield  {author} {\bibinfo {author} {\bibfnamefont {O.}~\bibnamefont
  {Entin-Wohlman}}, \bibinfo {author} {\bibfnamefont {Y.}~\bibnamefont
  {Gefen}},\ and\ \bibinfo {author} {\bibfnamefont {Y.}~\bibnamefont
  {Shapira}},\ }\bibfield  {title} {\bibinfo {title} {Variable-range hopping
  conductivity in granular materials},\ }\href
  {https://doi.org/10.1088/0022-3719/16/7/004} {\bibfield  {journal} {\bibinfo
  {journal} {Journal of Physics C: Solid State Physics}\ }\textbf {\bibinfo
  {volume} {16}},\ \bibinfo {pages} {1161} (\bibinfo {year}
  {1983})}\BibitemShut {NoStop}%
\bibitem [{\citenamefont {Lee}\ and\ \citenamefont
  {Ramakrishnan}(1985)}]{Lee:1985}%
  \BibitemOpen
  \bibfield  {author} {\bibinfo {author} {\bibfnamefont {P.}~\bibnamefont
  {Lee}}\ and\ \bibinfo {author} {\bibfnamefont {T.}~\bibnamefont
  {Ramakrishnan}},\ }\bibfield  {title} {\bibinfo {title} {Disordered
  electronic systems},\ }\href {https://doi.org/10.1103/RevModPhys.57.287}
  {\bibfield  {journal} {\bibinfo  {journal} {Rev.\ Mod.\ Phys.}\ }\textbf
  {\bibinfo {volume} {57}},\ \bibinfo {pages} {287} (\bibinfo {year}
  {1985})}\BibitemShut {NoStop}%
\bibitem [{\citenamefont {Thouless}(1977)}]{Thouless:1977}%
  \BibitemOpen
  \bibfield  {author} {\bibinfo {author} {\bibfnamefont {D.}~\bibnamefont
  {Thouless}},\ }\bibfield  {title} {\bibinfo {title} {Maximum metallic
  resistance in thin wires},\ }\href
  {https://doi.org/10.1103/PhysRevLett.39.1167} {\bibfield  {journal} {\bibinfo
   {journal} {Phys.\ Rev.\ Lett.}\ }\textbf {\bibinfo {volume} {39}},\ \bibinfo
  {pages} {1167} (\bibinfo {year} {1977})}\BibitemShut {NoStop}%
\bibitem [{\citenamefont {Wang}\ and\ \citenamefont
  {Santiago-ALves}(2006)}]{Wang:2006}%
  \BibitemOpen
  \bibfield  {author} {\bibinfo {author} {\bibfnamefont {Y.}~\bibnamefont
  {Wang}}\ and\ \bibinfo {author} {\bibfnamefont {J.}~\bibnamefont
  {Santiago-ALves}},\ }\bibfield  {title} {\bibinfo {title} {Large negative
  magnetoresistanca and strong localization in highly disordered elecrospun
  pregraphic carbon nanofiber},\ }\href {https://doi.org/10.1063/1.2338573}
  {\bibfield  {journal} {\bibinfo  {journal} {Appl.\ Phys.\ Lett.}\ }\textbf
  {\bibinfo {volume} {89}},\ \bibinfo {pages} {123119} (\bibinfo {year}
  {2006})}\BibitemShut {NoStop}%
\bibitem [{\citenamefont {Du}\ \emph {et~al.}(1998)\citenamefont {Du},
  \citenamefont {Prigodin}, \citenamefont {Burns}, \citenamefont {Joo},
  \citenamefont {Wang},\ and\ \citenamefont {Epstein}}]{Du:1998}%
  \BibitemOpen
  \bibfield  {author} {\bibinfo {author} {\bibfnamefont {G.}~\bibnamefont
  {Du}}, \bibinfo {author} {\bibfnamefont {V.}~\bibnamefont {Prigodin}},
  \bibinfo {author} {\bibfnamefont {A.}~\bibnamefont {Burns}}, \bibinfo
  {author} {\bibfnamefont {J.}~\bibnamefont {Joo}}, \bibinfo {author}
  {\bibfnamefont {C.}~\bibnamefont {Wang}},\ and\ \bibinfo {author}
  {\bibfnamefont {A.}~\bibnamefont {Epstein}},\ }\bibfield  {title} {\bibinfo
  {title} {Unusual semimetallic behavior of carbonized ion-implanted
  polymers},\ }\href {https://doi.org/10.1103/PhysRevB.58.4485} {\bibfield
  {journal} {\bibinfo  {journal} {Phys.\ Rev.\ B}\ }\textbf {\bibinfo {volume}
  {58}},\ \bibinfo {pages} {4485} (\bibinfo {year} {1998})}\BibitemShut
  {NoStop}%
\bibitem [{\citenamefont {Altshuler}\ and\ \citenamefont
  {Aronov}(1985)}]{Altshuler:1985}%
  \BibitemOpen
  \bibfield  {author} {\bibinfo {author} {\bibfnamefont {B.}~\bibnamefont
  {Altshuler}}\ and\ \bibinfo {author} {\bibfnamefont {A.}~\bibnamefont
  {Aronov}},\ }\bibinfo {title} {Electron-electron interactions in disordered
  systems}\ (\bibinfo  {publisher} {North-Holland, Amsterdam},\ \bibinfo {year}
  {1985})\ Chap.~\bibinfo {chapter} {1},\ \bibinfo {edition} {1st}\
  ed.\BibitemShut {Stop}%
\bibitem [{\citenamefont {Kobayashi}\ and\ \citenamefont
  {Komori}(1985)}]{Kobayashi:1985}%
  \BibitemOpen
  \bibfield  {author} {\bibinfo {author} {\bibfnamefont {S.}~\bibnamefont
  {Kobayashi}}\ and\ \bibinfo {author} {\bibfnamefont {F.}~\bibnamefont
  {Komori}},\ }\bibfield  {title} {\bibinfo {title} {Experiments on
  localization and interaction effects in metallic films},\ }\href
  {https://doi.org/10.1143/PTPS.84.224} {\bibfield  {journal} {\bibinfo
  {journal} {Prog.\ Theor.\ Phys.\ Suppl.}\ }\textbf {\bibinfo {volume} {84}},\
  \bibinfo {pages} {224} (\bibinfo {year} {1985})}\BibitemShut {NoStop}%
\bibitem [{\citenamefont {Sett}\ \emph {et~al.}(2017)\citenamefont {Sett},
  \citenamefont {Das},\ and\ \citenamefont {Raychaudhuri}}]{Sett_2017}%
  \BibitemOpen
  \bibfield  {author} {\bibinfo {author} {\bibfnamefont {S.}~\bibnamefont
  {Sett}}, \bibinfo {author} {\bibfnamefont {K.}~\bibnamefont {Das}},\ and\
  \bibinfo {author} {\bibfnamefont {A.~K.}\ \bibnamefont {Raychaudhuri}},\
  }\bibfield  {title} {\bibinfo {title} {Weak localization and the approach to
  metal-insulator transition in single crystalline germanium nanowires},\
  }\href {https://doi.org/10.1088/1361-648x/aa58fe} {\bibfield  {journal}
  {\bibinfo  {journal} {Journal of Physics: Condensed Matter}\ }\textbf
  {\bibinfo {volume} {29}},\ \bibinfo {pages} {115301} (\bibinfo {year}
  {2017})}\BibitemShut {NoStop}%
\bibitem [{\citenamefont {Hsu}\ \emph {et~al.}(1999)\citenamefont {Hsu},
  \citenamefont {Sheng},\ and\ \citenamefont {Lin}}]{PhysRevB.60.3940}%
  \BibitemOpen
  \bibfield  {author} {\bibinfo {author} {\bibfnamefont {S.~Y.}\ \bibnamefont
  {Hsu}}, \bibinfo {author} {\bibfnamefont {P.~J.}\ \bibnamefont {Sheng}},\
  and\ \bibinfo {author} {\bibfnamefont {J.~J.}\ \bibnamefont {Lin}},\
  }\bibfield  {title} {\bibinfo {title} {Quadratic temperature dependence of
  the electron-phonon scattering rate in disordered metals},\ }\href
  {https://doi.org/10.1103/PhysRevB.60.3940} {\bibfield  {journal} {\bibinfo
  {journal} {Phys. Rev. B}\ }\textbf {\bibinfo {volume} {60}},\ \bibinfo
  {pages} {3940} (\bibinfo {year} {1999})}\BibitemShut {NoStop}%
\bibitem [{\citenamefont {Lin}\ and\ \citenamefont {Bird}(2002)}]{Lin_2002}%
  \BibitemOpen
  \bibfield  {author} {\bibinfo {author} {\bibfnamefont {J.~J.}\ \bibnamefont
  {Lin}}\ and\ \bibinfo {author} {\bibfnamefont {J.~P.}\ \bibnamefont {Bird}},\
  }\bibfield  {title} {\bibinfo {title} {Recent experimental studies of
  electron dephasing in metal and semiconductor mesoscopic structures},\ }\href
  {https://doi.org/10.1088/0953-8984/14/18/201} {\bibfield  {journal} {\bibinfo
   {journal} {Journal of Physics: Condensed Matter}\ }\textbf {\bibinfo
  {volume} {14}},\ \bibinfo {pages} {R501} (\bibinfo {year}
  {2002})}\BibitemShut {NoStop}%
\bibitem [{\citenamefont {Ghosh}\ \emph {et~al.}(1998)\citenamefont {Ghosh},
  \citenamefont {Barman}, \citenamefont {De},\ and\ \citenamefont
  {Chatterjee}}]{Ghosh:1998}%
  \BibitemOpen
  \bibfield  {author} {\bibinfo {author} {\bibfnamefont {M.}~\bibnamefont
  {Ghosh}}, \bibinfo {author} {\bibfnamefont {A.}~\bibnamefont {Barman}},
  \bibinfo {author} {\bibfnamefont {S.}~\bibnamefont {De}},\ and\ \bibinfo
  {author} {\bibfnamefont {S.}~\bibnamefont {Chatterjee}},\ }\bibfield  {title}
  {\bibinfo {title} {Crossover from mott to efros-shklovskii
  variable-range-hopping conductivity in conducting polyaniline},\ }\href
  {https://doi.org/10.1016/S0379-6779(98)00105-2} {\bibfield  {journal}
  {\bibinfo  {journal} {Synthetic Metals}\ }\textbf {\bibinfo {volume} {97}},\
  \bibinfo {pages} {23} (\bibinfo {year} {1998})}\BibitemShut {NoStop}%
\bibitem [{\citenamefont {Zabrodskii}\ \emph {et~al.}(1998)\citenamefont
  {Zabrodskii}, \citenamefont {Andreev},\ and\ \citenamefont
  {Egorov}}]{Zabrodskii:1998}%
  \BibitemOpen
  \bibfield  {author} {\bibinfo {author} {\bibfnamefont {A.}~\bibnamefont
  {Zabrodskii}}, \bibinfo {author} {\bibfnamefont {A.}~\bibnamefont
  {Andreev}},\ and\ \bibinfo {author} {\bibfnamefont {S.}~\bibnamefont
  {Egorov}},\ }\bibfield  {title} {\bibinfo {title} {Coulomb gap and the
  metal-insulator transition},\ }\href
  {https://doi.org/10.1002/(SICI)1521-3951(199801)205:1<61::AID-PSSB61>3.0.CO;2-S}
  {\bibfield  {journal} {\bibinfo  {journal} {Phys.\ Stat.\ Sol.\ (b)}\
  }\textbf {\bibinfo {volume} {205}},\ \bibinfo {pages} {61} (\bibinfo {year}
  {1998})}\BibitemShut {NoStop}%
\bibitem [{\citenamefont {Hikami}\ \emph {et~al.}(1980)\citenamefont {Hikami},
  \citenamefont {Larkin},\ and\ \citenamefont {Nagaoka}}]{Hikami:1980}%
  \BibitemOpen
  \bibfield  {author} {\bibinfo {author} {\bibfnamefont {S.}~\bibnamefont
  {Hikami}}, \bibinfo {author} {\bibfnamefont {A.}~\bibnamefont {Larkin}},\
  and\ \bibinfo {author} {\bibfnamefont {Y.}~\bibnamefont {Nagaoka}},\
  }\bibfield  {title} {\bibinfo {title} {Spin-orbit interaction and
  magnetoresistance in the two dimensional random system},\ }\href
  {https://doi.org/10.1143/PTP.63.707} {\bibfield  {journal} {\bibinfo
  {journal} {Prog.\ Theor.\ Phys.}\ }\textbf {\bibinfo {volume} {63}},\
  \bibinfo {pages} {707} (\bibinfo {year} {1980})}\BibitemShut {NoStop}%
\bibitem [{\citenamefont {Bergmann}(1984)}]{Bergmann:1984}%
  \BibitemOpen
  \bibfield  {author} {\bibinfo {author} {\bibfnamefont {G.}~\bibnamefont
  {Bergmann}},\ }\bibfield  {title} {\bibinfo {title} {Weak localization in
  thin films: a time-of-flight experiment with conduction electrons},\ }\href
  {https://doi.org/10.1016/0370-1573(84)90103-0} {\bibfield  {journal}
  {\bibinfo  {journal} {Phys.\ Rep.}\ }\textbf {\bibinfo {volume} {107}},\
  \bibinfo {pages} {1} (\bibinfo {year} {1984})}\BibitemShut {NoStop}%
\bibitem [{\citenamefont {Akhgar}\ \emph {et~al.}(2016)\citenamefont {Akhgar},
  \citenamefont {Klochan}, \citenamefont {{Willems van Beveren}}, \citenamefont
  {Edmonds}, \citenamefont {Maier}, \citenamefont {Spencer}, \citenamefont
  {McCallum}, \citenamefont {Ley}, \citenamefont {Hamilton},\ and\
  \citenamefont {Pakes}}]{Akhgar:2016}%
  \BibitemOpen
  \bibfield  {author} {\bibinfo {author} {\bibfnamefont {G.}~\bibnamefont
  {Akhgar}}, \bibinfo {author} {\bibfnamefont {O.}~\bibnamefont {Klochan}},
  \bibinfo {author} {\bibfnamefont {L.}~\bibnamefont {{Willems van Beveren}}},
  \bibinfo {author} {\bibfnamefont {M.}~\bibnamefont {Edmonds}}, \bibinfo
  {author} {\bibfnamefont {F.}~\bibnamefont {Maier}}, \bibinfo {author}
  {\bibfnamefont {B.}~\bibnamefont {Spencer}}, \bibinfo {author} {\bibfnamefont
  {J.}~\bibnamefont {McCallum}}, \bibinfo {author} {\bibfnamefont
  {L.}~\bibnamefont {Ley}}, \bibinfo {author} {\bibfnamefont {A.}~\bibnamefont
  {Hamilton}},\ and\ \bibinfo {author} {\bibfnamefont {C.}~\bibnamefont
  {Pakes}},\ }\bibfield  {title} {\bibinfo {title} {Strong and tunable
  spin-orbit coupling in a two-dimensional hole gas in ionic-liquid gated
  diamond devices},\ }\href {https://doi.org/10.102/acs.nanolett.6b01155}
  {\bibfield  {journal} {\bibinfo  {journal} {Nano Lett.}\ }\textbf {\bibinfo
  {volume} {16}},\ \bibinfo {pages} {3768} (\bibinfo {year}
  {2016})}\BibitemShut {NoStop}%
\bibitem [{\citenamefont {Kawabata}(1980)}]{Kawabata:1980}%
  \BibitemOpen
  \bibfield  {author} {\bibinfo {author} {\bibfnamefont {A.}~\bibnamefont
  {Kawabata}},\ }\bibfield  {title} {\bibinfo {title} {Theory of negative
  magnetoresistance i. application to heavily doped semiconductors},\ }\href
  {https://doi.org/10.1143/JPSJ.49.628} {\bibfield  {journal} {\bibinfo
  {journal} {J.\ Phys.\ Soc.\ Jpn.}\ }\textbf {\bibinfo {volume} {49}},\
  \bibinfo {pages} {628} (\bibinfo {year} {1980})}\BibitemShut {NoStop}%
\bibitem [{\citenamefont {Bryksin}\ and\ \citenamefont
  {Kleinert}(1996)}]{BKtheory:1996}%
  \BibitemOpen
  \bibfield  {author} {\bibinfo {author} {\bibfnamefont {V.}~\bibnamefont
  {Bryksin}}\ and\ \bibinfo {author} {\bibfnamefont {P.}~\bibnamefont
  {Kleinert}},\ }\bibfield  {title} {\bibinfo {title} {Anderson localization in
  anisotropic systems at an arbitrary orientation of the magnetic field},\
  }\href {https://doi.org/10.1007/s002570050185} {\bibfield  {journal}
  {\bibinfo  {journal} {Z.\ Phys.\ B}\ }\textbf {\bibinfo {volume} {101}},\
  \bibinfo {pages} {91} (\bibinfo {year} {1996})}\BibitemShut {NoStop}%
\bibitem [{\citenamefont {Gougam}\ \emph {et~al.}(1999)\citenamefont {Gougam},
  \citenamefont {Gandit}, \citenamefont {Sicart},\ and\ \citenamefont
  {Robert}}]{Gougam:1999}%
  \BibitemOpen
  \bibfield  {author} {\bibinfo {author} {\bibfnamefont {A.}~\bibnamefont
  {Gougam}}, \bibinfo {author} {\bibfnamefont {P.}~\bibnamefont {Gandit}},
  \bibinfo {author} {\bibfnamefont {J.}~\bibnamefont {Sicart}},\ and\ \bibinfo
  {author} {\bibfnamefont {J.}~\bibnamefont {Robert}},\ }\bibfield  {title}
  {\bibinfo {title} {Negative magnetoresistance and electron localization in
  gaas-alas superlattices},\ }\href
  {https://doi.org/10.1088/0268-1242/14/3/005} {\bibfield  {journal} {\bibinfo
  {journal} {Semicond.\ Sci.\ Technol.}\ }\textbf {\bibinfo {volume} {14}},\
  \bibinfo {pages} {231} (\bibinfo {year} {1999})}\BibitemShut {NoStop}%
\bibitem [{\citenamefont {Bhattacharyya}\ and\ \citenamefont
  {Churochkin}(2014)}]{Bhattacharyya:2014}%
  \BibitemOpen
  \bibfield  {author} {\bibinfo {author} {\bibfnamefont {S.}~\bibnamefont
  {Bhattacharyya}}\ and\ \bibinfo {author} {\bibfnamefont {D.}~\bibnamefont
  {Churochkin}},\ }\bibfield  {title} {\bibinfo {title} {Polarization dependent
  asymmetric magneto-resistance features in nanocrystalline diamond films},\
  }\href {https://doi.org/10.1063/1.4893662} {\bibfield  {journal} {\bibinfo
  {journal} {Appl.\ Phys.\ Lett.}\ }\textbf {\bibinfo {volume} {105}},\
  \bibinfo {pages} {073111} (\bibinfo {year} {2014})}\BibitemShut {NoStop}%
\bibitem [{\citenamefont {Vlasov}\ \emph {et~al.}(2012)\citenamefont {Vlasov},
  \citenamefont {Kanzyuba}, \citenamefont {Shiryaev}, \citenamefont {Volkov},
  \citenamefont {Ralchenko},\ and\ \citenamefont {Konov}}]{Vlasov:2012}%
  \BibitemOpen
  \bibfield  {author} {\bibinfo {author} {\bibfnamefont {I.}~\bibnamefont
  {Vlasov}}, \bibinfo {author} {\bibfnamefont {M.}~\bibnamefont {Kanzyuba}},
  \bibinfo {author} {\bibfnamefont {A.}~\bibnamefont {Shiryaev}}, \bibinfo
  {author} {\bibfnamefont {V.}~\bibnamefont {Volkov}}, \bibinfo {author}
  {\bibfnamefont {V.}~\bibnamefont {Ralchenko}},\ and\ \bibinfo {author}
  {\bibfnamefont {V.}~\bibnamefont {Konov}},\ }\bibfield  {title} {\bibinfo
  {title} {Percolation model of an insulator-conductor transition in
  ultrananocrystalline diamond fils},\ }\href
  {https://doi.org/10.1134/S0021364012070090} {\bibfield  {journal} {\bibinfo
  {journal} {JETP.\ Lett.}\ }\textbf {\bibinfo {volume} {95}},\ \bibinfo
  {pages} {391} (\bibinfo {year} {2012})}\BibitemShut {NoStop}%
\bibitem [{\citenamefont {Hejazi}\ \emph {et~al.}(2019)\citenamefont {Hejazi},
  \citenamefont {Tong}, \citenamefont {Stacey}, \citenamefont {Soto-Breceda},
  \citenamefont {Ibbotson}, \citenamefont {Yunzab}, \citenamefont {Maturana},
  \citenamefont {Almasi}, \citenamefont {Jung}, \citenamefont {Sun},
  \citenamefont {Meffin}, \citenamefont {Fang}, \citenamefont {Stamp},
  \citenamefont {Ganesan}, \citenamefont {Fox}, \citenamefont {Rifai},
  \citenamefont {Nadarajah}, \citenamefont {Falahatdoost}, \citenamefont
  {Prawer}, \citenamefont {Apollo},\ and\ \citenamefont
  {Garrett}}]{Hejazi:2019aa}%
  \BibitemOpen
  \bibfield  {author} {\bibinfo {author} {\bibfnamefont {M.~A.}\ \bibnamefont
  {Hejazi}}, \bibinfo {author} {\bibfnamefont {W.}~\bibnamefont {Tong}},
  \bibinfo {author} {\bibfnamefont {A.}~\bibnamefont {Stacey}}, \bibinfo
  {author} {\bibfnamefont {A.}~\bibnamefont {Soto-Breceda}}, \bibinfo {author}
  {\bibfnamefont {M.~R.}\ \bibnamefont {Ibbotson}}, \bibinfo {author}
  {\bibfnamefont {M.}~\bibnamefont {Yunzab}}, \bibinfo {author} {\bibfnamefont
  {M.~I.}\ \bibnamefont {Maturana}}, \bibinfo {author} {\bibfnamefont
  {A.}~\bibnamefont {Almasi}}, \bibinfo {author} {\bibfnamefont {Y.~J.}\
  \bibnamefont {Jung}}, \bibinfo {author} {\bibfnamefont {S.}~\bibnamefont
  {Sun}}, \bibinfo {author} {\bibfnamefont {H.}~\bibnamefont {Meffin}},
  \bibinfo {author} {\bibfnamefont {J.}~\bibnamefont {Fang}}, \bibinfo {author}
  {\bibfnamefont {M.~E.~M.}\ \bibnamefont {Stamp}}, \bibinfo {author}
  {\bibfnamefont {K.}~\bibnamefont {Ganesan}}, \bibinfo {author} {\bibfnamefont
  {K.}~\bibnamefont {Fox}}, \bibinfo {author} {\bibfnamefont {A.}~\bibnamefont
  {Rifai}}, \bibinfo {author} {\bibfnamefont {A.}~\bibnamefont {Nadarajah}},
  \bibinfo {author} {\bibfnamefont {S.}~\bibnamefont {Falahatdoost}}, \bibinfo
  {author} {\bibfnamefont {S.}~\bibnamefont {Prawer}}, \bibinfo {author}
  {\bibfnamefont {N.~V.}\ \bibnamefont {Apollo}},\ and\ \bibinfo {author}
  {\bibfnamefont {D.~J.}\ \bibnamefont {Garrett}},\ }\bibfield  {title}
  {\bibinfo {title} {Hybrid diamond/carbon fiber microelectrodes enable
  multimodal electrical/chemical neural interfacing},\ }\href
  {https://doi.org/https://doi.org/10.1016/j.biomaterials.2019.119648}
  {\bibfield  {journal} {\bibinfo  {journal} {Biomaterials}\ ,\ \bibinfo
  {pages} {119648}} (\bibinfo {year} {2019})}\BibitemShut {NoStop}%
\bibitem [{\citenamefont {Chimowa}\ \emph {et~al.}(2012)\citenamefont
  {Chimowa}, \citenamefont {Churochkin},\ and\ \citenamefont
  {Bhattacharyya}}]{Chimowa:2012}%
  \BibitemOpen
  \bibfield  {author} {\bibinfo {author} {\bibfnamefont {G.}~\bibnamefont
  {Chimowa}}, \bibinfo {author} {\bibfnamefont {D.}~\bibnamefont
  {Churochkin}},\ and\ \bibinfo {author} {\bibfnamefont {S.}~\bibnamefont
  {Bhattacharyya}},\ }\bibfield  {title} {\bibinfo {title} {Conductivity
  crossover in nano-crystalline diamond films: Realization of a disordered
  superlattice-like structure},\ }\href
  {https://doi.org/10.1209/0295-5075/99/27004} {\bibfield  {journal} {\bibinfo
  {journal} {EPL}\ }\textbf {\bibinfo {volume} {99}},\ \bibinfo {pages} {27004}
  (\bibinfo {year} {2012})}\BibitemShut {NoStop}%
\bibitem [{\citenamefont {Pusep}\ \emph {et~al.}(2005)\citenamefont {Pusep},
  \citenamefont {Ribeiro}, \citenamefont {Arakaki}, \citenamefont {de~Souza},
  \citenamefont {Malzer},\ and\ \citenamefont {D$\ddot{o}$hler}}]{Pusep:2005}%
  \BibitemOpen
  \bibfield  {author} {\bibinfo {author} {\bibfnamefont {Y.}~\bibnamefont
  {Pusep}}, \bibinfo {author} {\bibfnamefont {M.}~\bibnamefont {Ribeiro}},
  \bibinfo {author} {\bibfnamefont {H.}~\bibnamefont {Arakaki}}, \bibinfo
  {author} {\bibfnamefont {C.}~\bibnamefont {de~Souza}}, \bibinfo {author}
  {\bibfnamefont {S.}~\bibnamefont {Malzer}},\ and\ \bibinfo {author}
  {\bibfnamefont {G.}~\bibnamefont {D$\ddot{o}$hler}},\ }\bibfield  {title}
  {\bibinfo {title} {Disorder-driven coherence-incoherence crossover in random
  gaas/al$_{0.3}$ga$_{0.7}$as superlattices},\ }\href
  {https://doi.org/10.1103/PhysRevB.71.035323} {\bibfield  {journal} {\bibinfo
  {journal} {Phys.\ Rev.\ B}\ }\textbf {\bibinfo {volume} {71}},\ \bibinfo
  {pages} {035323} (\bibinfo {year} {2005})}\BibitemShut {NoStop}%
\bibitem [{\citenamefont {Chiquito}\ \emph {et~al.}(2002)\citenamefont
  {Chiquito}, \citenamefont {Pusep}, \citenamefont {Gusev}, \citenamefont
  {Gusev},\ and\ \citenamefont {Toropov}}]{Chiquito:2002}%
  \BibitemOpen
  \bibfield  {author} {\bibinfo {author} {\bibfnamefont {A.}~\bibnamefont
  {Chiquito}}, \bibinfo {author} {\bibfnamefont {Y.}~\bibnamefont {Pusep}},
  \bibinfo {author} {\bibfnamefont {G.}~\bibnamefont {Gusev}}, \bibinfo
  {author} {\bibfnamefont {G.}~\bibnamefont {Gusev}},\ and\ \bibinfo {author}
  {\bibfnamefont {A.}~\bibnamefont {Toropov}},\ }\bibfield  {title} {\bibinfo
  {title} {Quantum interference in intentionally disordered
  gaas/al$_{x}$ga$_{1-x}$as superlattices},\ }\href
  {https://doi.org/10.1103/PhysRevB.66.035323} {\bibfield  {journal} {\bibinfo
  {journal} {Phys.\ Rev.\ B}\ }\textbf {\bibinfo {volume} {66}},\ \bibinfo
  {pages} {035323} (\bibinfo {year} {2002})}\BibitemShut {NoStop}%
\bibitem [{\citenamefont {Pusep}\ \emph {et~al.}(2003)\citenamefont {Pusep},
  \citenamefont {Ribeiro}, \citenamefont {Arakaki}, \citenamefont {de~Souza},
  \citenamefont {Zanello}, \citenamefont {Chiquito}, \citenamefont {Malzer},\
  and\ \citenamefont {D$\ddot{o}$hler}}]{Pusep:2003}%
  \BibitemOpen
  \bibfield  {author} {\bibinfo {author} {\bibfnamefont {Y.}~\bibnamefont
  {Pusep}}, \bibinfo {author} {\bibfnamefont {M.}~\bibnamefont {Ribeiro}},
  \bibinfo {author} {\bibfnamefont {H.}~\bibnamefont {Arakaki}}, \bibinfo
  {author} {\bibfnamefont {C.}~\bibnamefont {de~Souza}}, \bibinfo {author}
  {\bibfnamefont {P.}~\bibnamefont {Zanello}}, \bibinfo {author} {\bibfnamefont
  {A.}~\bibnamefont {Chiquito}}, \bibinfo {author} {\bibfnamefont
  {S.}~\bibnamefont {Malzer}},\ and\ \bibinfo {author} {\bibfnamefont
  {G.}~\bibnamefont {D$\ddot{o}$hler}},\ }\bibfield  {title} {\bibinfo {title}
  {Anisotropy of quantum interference in disordered gaas/al$_{x}$ga$_{1-x}$as
  superlattices},\ }\href {https://doi.org/10.1103/PhysRevB.68.195207}
  {\bibfield  {journal} {\bibinfo  {journal} {Phys.\ Rev.\ B}\ }\textbf
  {\bibinfo {volume} {68}},\ \bibinfo {pages} {195207} (\bibinfo {year}
  {2003})}\BibitemShut {NoStop}%
\bibitem [{\citenamefont {Chassam-Chenai}\ and\ \citenamefont
  {Mailly}(1995)}]{Chenai:1995}%
  \BibitemOpen
  \bibfield  {author} {\bibinfo {author} {\bibfnamefont {A.}~\bibnamefont
  {Chassam-Chenai}}\ and\ \bibinfo {author} {\bibfnamefont {D.}~\bibnamefont
  {Mailly}},\ }\bibfield  {title} {\bibinfo {title} {Transport in
  quasi-two-dimensional systems under a weak magnetic field},\ }\href
  {https://doi.org/10.1103/PhysRevB.52.1984} {\bibfield  {journal} {\bibinfo
  {journal} {Phys.\ Rev.\ B}\ }\textbf {\bibinfo {volume} {52}},\ \bibinfo
  {pages} {1984} (\bibinfo {year} {1995})}\BibitemShut {NoStop}%
\bibitem [{\citenamefont {Sagar}\ \emph {et~al.}(2015)\citenamefont {Sagar},
  \citenamefont {Saleemi},\ and\ \citenamefont {Zhang}}]{Sagar:2015a}%
  \BibitemOpen
  \bibfield  {author} {\bibinfo {author} {\bibfnamefont {R.}~\bibnamefont
  {Sagar}}, \bibinfo {author} {\bibfnamefont {A.}~\bibnamefont {Saleemi}},\
  and\ \bibinfo {author} {\bibfnamefont {X.}~\bibnamefont {Zhang}},\ }\bibfield
   {title} {\bibinfo {title} {Angular magnetoresistance in semiconducting
  undoped amorphous carbon thin films},\ }\href
  {https://doi.org/10.1063/1.4919820} {\bibfield  {journal} {\bibinfo
  {journal} {J.\ Appl.\ Phys.}\ }\textbf {\bibinfo {volume} {117}},\ \bibinfo
  {pages} {174503} (\bibinfo {year} {2015})}\BibitemShut {NoStop}%
\bibitem [{\citenamefont {Sagar}\ \emph {et~al.}(2017)\citenamefont {Sagar},
  \citenamefont {Saleemi}, \citenamefont {Shehzad}, \citenamefont {Navale},
  \citenamefont {Mane},\ and\ \citenamefont {Stadler}}]{Sagar:2015b}%
  \BibitemOpen
  \bibfield  {author} {\bibinfo {author} {\bibfnamefont {R.}~\bibnamefont
  {Sagar}}, \bibinfo {author} {\bibfnamefont {A.}~\bibnamefont {Saleemi}},
  \bibinfo {author} {\bibfnamefont {K.}~\bibnamefont {Shehzad}}, \bibinfo
  {author} {\bibfnamefont {S.}~\bibnamefont {Navale}}, \bibinfo {author}
  {\bibfnamefont {R.}~\bibnamefont {Mane}},\ and\ \bibinfo {author}
  {\bibfnamefont {F.}~\bibnamefont {Stadler}},\ }\bibfield  {title} {\bibinfo
  {title} {Non-magnetic thin films for magnetic field position sensor},\ }\href
  {https://doi.org/10.1016/j.sna.2016.12.001} {\bibfield  {journal} {\bibinfo
  {journal} {Sens.\ Actuators A:\ Phys.}\ }\textbf {\bibinfo {volume} {254}},\
  \bibinfo {pages} {89} (\bibinfo {year} {2017})}\BibitemShut {NoStop}%
\bibitem [{\citenamefont {Yuan}\ \emph {et~al.}(2016)\citenamefont {Yuan},
  \citenamefont {Fang}, \citenamefont {Feng}, \citenamefont {Chen},
  \citenamefont {Wen}, \citenamefont {Xiong},\ and\ \citenamefont
  {Wang}}]{Yuan:2016}%
  \BibitemOpen
  \bibfield  {author} {\bibinfo {author} {\bibfnamefont {W.}~\bibnamefont
  {Yuan}}, \bibinfo {author} {\bibfnamefont {L.}~\bibnamefont {Fang}}, \bibinfo
  {author} {\bibfnamefont {Z.}~\bibnamefont {Feng}}, \bibinfo {author}
  {\bibfnamefont {Z.}~\bibnamefont {Chen}}, \bibinfo {author} {\bibfnamefont
  {J.}~\bibnamefont {Wen}}, \bibinfo {author} {\bibfnamefont {Y.}~\bibnamefont
  {Xiong}},\ and\ \bibinfo {author} {\bibfnamefont {B.}~\bibnamefont {Wang}},\
  }\bibfield  {title} {\bibinfo {title} {Highly conductive nitrogen-doped
  ultrananocrystalline diamond films with enhanced field emission properties:
  triethylamine as a new nitrogen source},\ }\href
  {https://doi.org/10.1039/c6tc00087h} {\bibfield  {journal} {\bibinfo
  {journal} {J.\ Mater.\ Chem.\ C}\ }\textbf {\bibinfo {volume} {4}},\ \bibinfo
  {pages} {4778} (\bibinfo {year} {2016})}\BibitemShut {NoStop}%
\bibitem [{\citenamefont {Ganesan}\ \emph {et~al.}(2014)\citenamefont
  {Ganesan}, \citenamefont {Garrett}, \citenamefont {Ahnood}, \citenamefont
  {Shivdasani}, \citenamefont {Tong}, \citenamefont {Turnley}, \citenamefont
  {Fox}, \citenamefont {Meffin},\ and\ \citenamefont {Prawer}}]{Ganesan:2014}%
  \BibitemOpen
  \bibfield  {author} {\bibinfo {author} {\bibfnamefont {K.}~\bibnamefont
  {Ganesan}}, \bibinfo {author} {\bibfnamefont {D.}~\bibnamefont {Garrett}},
  \bibinfo {author} {\bibfnamefont {A.}~\bibnamefont {Ahnood}}, \bibinfo
  {author} {\bibfnamefont {M.}~\bibnamefont {Shivdasani}}, \bibinfo {author}
  {\bibfnamefont {W.}~\bibnamefont {Tong}}, \bibinfo {author} {\bibfnamefont
  {A.}~\bibnamefont {Turnley}}, \bibinfo {author} {\bibfnamefont
  {K.}~\bibnamefont {Fox}}, \bibinfo {author} {\bibfnamefont {H.}~\bibnamefont
  {Meffin}},\ and\ \bibinfo {author} {\bibfnamefont {S.}~\bibnamefont
  {Prawer}},\ }\bibfield  {title} {\bibinfo {title} {An all-diamond, hermetic
  electrical feedthrough array for a retinal prosthesis},\ }\href
  {https://doi.org/10.1016/j.biomaterials.2013.10.040} {\bibfield  {journal}
  {\bibinfo  {journal} {Biomaterials}\ }\textbf {\bibinfo {volume} {35}},\
  \bibinfo {pages} {908} (\bibinfo {year} {2014})}\BibitemShut {NoStop}%
\end{thebibliography}%

\end{document}